\documentclass[12pt,reqno]{amsart}
\allowdisplaybreaks[4]
\usepackage{graphicx} %We can use any other package if it is necessary
\usepackage{amssymb}
\usepackage{amsmath}
\usepackage{cite}

\newcommand{\be}{\begin{equation}}
\newcommand{\ee}{\end{equation}}
\newcommand{\bea}{\begin{eqnarray}}
\newcommand{\eea}{\end{eqnarray}}
\newcommand{\ba}{\begin{array}}
\newcommand{\ea}{\end{array}}
\newcommand{\bean}{\begin{eqnarray*}}
\newcommand{\eean}{\end{eqnarray*}}

\newcommand{\pa}{\partial}

\begin{document}
\title[Darboux]{ The Darboux transformation of
 the derivative nonlinear Schr\"odinger equation}
\author{Shuwei Xu\dag,Jingsong He$^*$ \dag,
 Lihong Wang \dag}

\thanks{$^*$ Corresponding author: hejingsong@nbu.edu.cn,jshe@ustc.edu.cn}

 \maketitle%%%%%%%%%%%%%%(Õâ¸öÃüÁî·ÅµÄλÖúÜÖØÒª)
\dedicatory {  \dag \ Department of Mathematics, Ningbo University,
Ningbo , Zhejiang 315211, P.\ R.\ China\\}

\begin{abstract}
The n-fold Darboux transformation (DT) is a 2¡Á2 matrix for the Kaup-Newell (KN) system. In this paper,
 each element of this matrix is expressed by a ratio of $(n+1)\times (n+1)$ determinant and
  $n\times n$ determinant of eigenfunctions. Using these formulae, the expressions of
   the $q^{[n]}$ and $r^{[n]}$ in KN system are generated by n-fold DT.
Further, under the reduction condition, the rogue wave,rational traveling solution, dark soliton, bright soliton,
breather solution, periodic solution of the derivative nonlinear Schr\"odinger(DNLS) equation
are given explicitly by different seed solutions. In particular, the rogue wave and
rational traveling solution are  two kinds of new solutions. The complete classification of these solutions
generated by one-fold DT is given in the table on page.\\
\end{abstract}

{\bf Key words}: derivative nonlinear Schr\"odinger equation,
Darboux transformation,  \\
\mbox{\hspace{3cm}} soliton, rational solution,  breather solution, rogue wave.

{\bf PACS(2010) numbers}: 02.30.Ik, 42.81.Dp, 52.35.Bj, 52.35.Sb, 94.05.Fg

{\bf MSC(2010) numbers}: 35C08, 37K10, 37K40

%35C08 Soliton solutions
%37K10 Completely integrable systems, integrability tests, bi-Hamiltonian structures, hierarchies (KdV, KP, Toda, etc.
%37K40 Soliton theory, asymptotic behavior of solutions

%%%02.30.Ik., integrable systems;
%% 42.81.Dp,Propagation, scattering, and losses; solitons
%52.35.Bj Magnetohydrodynamic waves (e.g., Alfven waves)
%52.35.Sb Solitons; BGK modes
%05.45.Yv Solitons (see 52.35.Sb for solitons in plasma;
%for solitons in acoustics,see 43.25.Rq¡ªin Acoustics Appendix;
%see 42.50.Md, 42.65.Tg, 42.81.Dp for solitons in optics;see also 03.75.Lm in matter waves;
%for solitons in space plasma physics, see 94.05.Fg;
%for solitary waves in fluid dynamics, see 47.35.Fg)

%%%%%%%%%%%%%%%%%%%%%%%%%%%%%%%%%%%%%%%%%%%%%%%%%%%%%%
\section{Introduction}
%%%%%%%%%%%%%%%%%%%%%%%%%%%%%%%%%%%%%%%%%%%%%%%%%%%%%%%
The derivative nonlinear Schr\"odinger equation,
\begin{equation}\label{dnlsI}
iq_{t}-q_{xx}+i(q^2q^\ast)_{x}=0,
\end{equation}
 one of the most important integrable systems in the mathematics and physics, is usually
called DNLS(or DNLSI) equation.  Here``*" denotes the complex
conjugation, and subscript of $x$ (or $t$) denotes the partial
derivative with respect to $x$ (or $t$).   This equation  is
originated from two fields of applied physics. The first is plasma
physics in which the DNLS governs the evolution of small but finite
amplitude Alfv\'en waves that propagates quasi-parallel to the
magnetic field\cite{MDNLS,MHD1}. Recently, this equation is also
used to describe large-amplitude magnetohydrodynamic (MHD) waves in
plasmas\cite{ruderman1,ruderman2}. Further, it is natural to improve
DNLS equation in more practical plasmas. For example,  DNLS
truncation model\cite{DNLStruncation} and the DNLS  with nonlinear
Landau damping\cite{landaudamping}.  In the second area, nonlinear
optics, the sub-picosecond or femtosecond pulses in single-mode
optical fiber is modeled by the DNLS \cite{WDNLS,anderson,govind}.

However, the crucial feature of the DNLS is that the integrability such as the dynamical evolution
of the associated physical system can be given analytically by using its exact solution.
Under the vanishing boundary condition(VBC), Kaup and Newell(KN)\cite{KN} firstly proposed an inverse
scattering transform (IST) with a revision in their pioneer works, and got a one-soliton solution.
Later, Kawata \cite{kawata1} further  solved DNLS under VBC and non-vanishing
boundary condition (NVBC) to get two soliton solution, and introduced "paired soliton" which is now
regarded as one kind of breather solution.  N-soliton formula \cite{huang1} of the DNLS with VBC is
expressed by determinants with the help of pole-expansion. Further, the IST of the DNLS with VBC is
re-considered by Huang's group\cite{huang2, huang3,huang4,huang5} and then the explicit form of the
N-soliton is obtained by some algebraic techniques. Now we turn to the DNLS under NVBC, and  some
special solutions are obtained and the existence  of the algebraic soliton is also given
\cite{ichikawa}.  This is followed by paired-soliton of  the DNLS from the IST \cite{kawata2}.
Wadati etal.\cite{wadati2} have given the  stationary solutions of the DNLS under the plane wave boundary
and the contributions of the derivative term in the DNLS equation. Recently, to avoid the multi-value problem,
Chen and Lam \cite{cheng1} revised the IST for the DNLS  under NVBC by  introducing an affine
parameter, and then got single breather solution, which can be reduced to the dark
soltion and bright soliton. Further applications on this method can be found in reference
\cite{lashkin1}. Cai and Huang\cite{huang6} found the action-angle variables of the DNLS
explicitly by constructing its Hamiltonian formalism.

Similar to many usual soliton equations, the DNLS is also solved by the Hirota method \cite{kakei1}
and Darboux transformation(DT) \cite{kenji1,steduel} besides IST.
By comparing with the corresponding results\cite{GN,matveev, he} of nonlinear Schr\"odinger(NLS) equation,
the DT \cite{kenji1,steduel} of the DNLS has following essential distinctness:
\begin{itemize}
\item  the kernel of one-fold DT is one-dimensional and it can be defined
  by one eigenfunction of linear system  defined by spectral problem,
\item the DNLS will be invariant under one-fold DT associated with a pure imaginary
eigenvalue(see the last paragraph of the section 2).
\end{itemize}
Some solutions\cite{kenji1} including multi-soliton and quasi-periodic solutions are obtained by
this DT from a trivial seed: zero solution(or vacuum). Steudel \cite{steduel} has obtained a general
formula of solutions $q^{[N]}$ and $r^{[N]}$ of KN system in terms of Vandermonde-like determinants by
N-fold DTs, and then given n-soliton and N-phase solutions from zero seed, N-breather solutions from non-zero seed:
monochromatic wave. Unlike the usual DT, Steduel used solutions of Riccati
equations, which are transformed from the linear partial differential equations of the spectral problem for
 the DNLS, to construct the solutions of the DNLS.
So the first difficulty of his method is to solve nonlinear Riccati, which is not solvable in general.
To overcome this difficulty, Steudel have made an Ansatz(see eq.(51) in reference \cite{steduel})
and introduced his favorite Seahorse functions. Moreover, the classification of the solutions(see
Figure 1 in reference\cite{steduel}) generated by DT is very interesting and useful. But the
conditions of parameters to generate dark soliton and bright soliton of the DNLS are not clear.
Therefore, it is natural to question whether the difficult Riccati equations  are  indeed
unavoidable for the DT from non-zero seeds and whether the classification of solutions generated by
one-fold DT can be fixed thoroughly or not.

It is interesting that the Ablowitz-Kaup-Newell-Segur(AKNS)system\cite{AKNS} can be mapped to the KN
system by a gauge transformation\cite{wadati1}.  Moreover, there exists other two kinds of derivative
nonlinear Schr\"odinger equation, i.e,.
the DNLSII \cite{CL}
\begin{equation}\label{CL}
iq_{t}+q_{xx}+iqq^\ast q_{x}=0,
\end{equation}
and the DNLSIII\cite{GI}
\begin{equation}\label{GI}
iq_{t}+q_{xx}-iq^2q^{\ast}_{x}+\frac{1}{2}q^3{q^\ast}^2=0,
\end{equation}
and a chain of gauge transformations between them:
DNLSII $\stackrel{a)}{\Longrightarrow}$
 DNLSI $\stackrel{b)}{\Longrightarrow}$ DNLSIII.
Here a) denotes eq.(2.12) in ref.\cite{wadati1}, and b) denotes: eq.(4)$\rightarrow$ eq.(3)
$\rightarrow$eq.(6) with $\gamma=0$ in ref.\cite{kakei1}. But these transformations can not preserve
the reduction conditions in spectral problem of the KN system and involve complicated integrations.
So each of them deserves investigating separately.

There are two aims of this paper. First aim is to present a detailed
derivation of the DT for the DNLS and  its determinant
representation. Using this representation, the solutions of DNLS can
be expressed by the solutions(eigenfunctions) of the linear partial
differential equations of the spectral problem of the KN system
instead of the solutions of the nonlinear Riccati equations, which
shows that the nonlinear Riccati equation and Seahorse functions are
indeed avoidable for the DT from nonzero seeds. A second aim is to
present a complete classification of the solutions generated by
one-fold DT from zero seed, non-zero seeds: constant solution and
periodic solution with a constant amplitude.

The organization of this paper is as follows. In section 2, it provides a relatively simple
approach to DT for the KN system, and then the determinant representation of the n-fold
DT  and  formulae of $q^{[n]}$ and $r^{[n]}$ expressed by eigenfunctions of spectral problem
are given. The reduction of DT of the KN system to the DNLS equation is also discussed by choosing
paired eigenvalues and eigenfunctions. In section 3, under specific reduction conditions, several
types of particular solutions are given from zero seed, non-zero seeds: constant solution and periodic
solution with a constant amplitude.  The complete classification of dark soliton, bright soliton,
periodic solution are given in a table for one-fold DT of the DNLS equation. In particular, two kinds of
new solutions: rational traveling solution and rogue wave are given.
The conclusion will be given in section 4.
%%%%%%%%%%%%%%%%%%%%%%%%%%%%%%%%%%%%%%%%%%%%%
\section{Darboux transformation}

Let us start from the first non-trivial flow of the KN
system\cite{KN},
\begin{equation}\label{sy1}
r_{t}-ir_{xx}-(r^2q)_{x}=0,
\end{equation}
\begin{equation}\label{sy2}
q_{t}+iq_{xx}-(rq^2)_{x}=0,
\end{equation}
which are  exactly reduced to the DNLS eq.(\ref{dnlsI}) for $r=-q^\ast$ while the choice $r
=q^\ast$ would lead to eq.(\ref{dnlsI}) with the sign of the nonlinear term changed. The Lax pairs
corresponding to coupled DNLS equations(\ref{sy1}) and (\ref{sy2}) can
be given by the Kaup¨CNewell spectral problem\cite{KN}
\begin{equation}\label{sys11}
 \pa_{x}\psi=(J\lambda^2+Q\lambda)\psi=U\psi,
\end{equation}
\begin{equation}\label{sys22}
\pa_{t}\psi=(2J\lambda^4+V_{3}\lambda^3+V_{2}\lambda^2+V_{1}\lambda)\psi=V\psi,
\end{equation}
with
\begin{equation}\label{fj1}
    \psi=\left( \begin{array}{c}
      \phi \\
      \varphi\\
     \end{array} \right),\nonumber\\
   \quad J= \left( \begin{array}{cc}
      i &0 \\
      0 &-i\\
   \end{array} \right),\nonumber\\
  \quad Q=\left( \begin{array}{cc}
     0 &q \\
     r &0\\
  \end{array} \right),\nonumber\\
\end{equation}
\begin{equation}\label{fj2}
V_{3}=2Q,  \quad V_{2}=Jqr, \quad V_{1}=\left( \begin{array}{cc}
0 &-iq_{x}+q^2r \\
 ir_{x}+r^2q&0\\
\end{array} \right).\nonumber\\
\end{equation}
Here $\lambda$, an arbitrary complex number, is  called the eigenvalue(or spectral parameter),
and $\psi$ is called the eigenfunction associated with  $\lambda$ of the KN system.
Equations(\ref{sy1}) and (\ref{sy2}) are equivalent to the integrability condition  $U_{t}-V_{x}+[U,V]=0$ of
(\ref{sys11}) and (\ref{sys22}).

The main task of this section is to present a detailed derivation of the Darboux transforation of the
DNLS and the determinant representation of the n-fold transformation.
Based on the DT for the NLS\cite{GN,matveev,he} and the DNLS\cite{kenji1,steduel},
 the main steps are : 1) to find a $2\times 2$ matrix  $T$ so that the KN spectral problem
 eq.(\ref{sys11}) and eq.(\ref{sys22}) is covariant, then get new
 solution $(q^{[1]},r^{[1]})$ expressed by elements of $T$ and
 seed solution $(q,r)$; 2) to find expressions of elements of $T$ in terms of eigenfunctions of KN spectral problem
corresponding to the seed solution $(q,r)$; 3) to get the determinant representation of n-fold DT $T_n $ and
new solutions $(q^{[n]},r^{[n]})$ by  $n$-times iteration of the DT;
4) to consider the reduction condition: $q^{[n]}= -(r^{[n]})^*$ by choosing special eigenvalue $\lambda_k$
 and its eigenfunction $\psi_k$, and then get $q^{[n]}$ of the DNLS equation expressed by its seed solution
 $q$ and its associated eigenfunctions $\{\psi_k,k=1,2,\cdots,n\}$. However, we shall use the kernel of n-fold DT($T_n$) to
fix it in the third step instead of iteration.

It is easy to see that the spectral problem (\ref{sys11}) and (\ref{sys22}) are transformed
to
  \begin{equation}\label{bh1}
{\psi^{[1]}}_{x}=U^{[1]}~\psi^{[1]},\ \ U^{[1]}=(T_{x}+T~U)T^{-1}.
\end{equation}
\begin{equation}\label{bh2}
{\psi^{[1]}}_{t}=V^{[1]}~\psi^{[1]}, \ \ V^{[1]}=(T_{t}+T~V)T^{-1}.
\end{equation}\\
under a gauge transformation \begin{equation}\label{bh3}
\psi^{[1]}=T~\psi.
\end{equation}
By cross differentiating (\ref{bh1}) and (\ref{bh2}), we obtain
\begin{equation}\label{bh4}
{U^{[1]}}_{t}-{V^{[1]}}_{x}+[{U^{[1]}},{V^{[1]}}]=T(U_{t}-V_{x}+[U,V])T^{-1}.
\end{equation}
This implies that, in order to make eqs.(\ref{sy1}) and eq.(\ref{sy2}) invariant under the
transformation (\ref{bh3}), it is crucial to search a matrix $T$ so that  $U^{[1]}$, $V^{[1]}$have
the same forms as $U$, $V$. At the same time the old potential(or
seed solution)($q$, $r$) in spectral matrixes $U$, $V$ are mapped
into new potentials
(or new solution)($q^{[1]}$, $r^{[1]}$) in transformed spectral matrixes $U^{[1]}$, $V^{[1]}$.\\

2.1 One-fold Darboux transformation of the KN system

Considering the universality of DT, suppose that the trial Darboux matrix $T$ in eq.(\ref{bh3}) is of
form
\begin{equation}\label{tt1}
T=T(\lambda)=\left( \begin{array}{cc}
a_{1}&b_{1} \\
c_{1} &d_{1}\\
\end{array} \right)\lambda+\left( \begin{array}{cc}
a_{0}&b_{0}\\
c_{0} &d_{0}\\
\end{array} \right),
\end{equation}
where $a_{0},  b_{0},  c_{0},   d_{0},   a_{1},   b_{1},   c_{1},   d_{1}$ are functions of $x$, $t$
to need be determined. From \begin{equation}\label{tt2} T_{x}+T~U=U^{[1]}~T,
\end{equation}
comparing the coefficients of $\lambda^{j}, j=3, 2, 1, 0$, it yields
\begin{eqnarray}\label{xx1}
&&\lambda^{3}: b_{1}=0,\ c_{1}=0,\nonumber\\
&&\lambda^{2}: q~a_{1}-2~i~b_{0}-q^{[1]}d_{1}=0,\ -r^{[1]}~a_{1}+r~d_{1}+2~i~c_{0}=0,\nonumber\\
&&\lambda^{1}: {a_{1}}_{x}+r~b_{0}-q^{[1]}c_{0}=0,\ {d_{1}}_{x}+q c_{0}-r^{[1]}b_{0}=0,
q a_{0}-q^{[1]}d_{0}=0,-r^{[1]}a_{0}+r d_{0}=0,\nonumber\\
&&\lambda^{0}: {a_{0}}_{x}={b_{0}}_{x}={c_{0}}_{x}={d_{0}}_{x}=0.
\end{eqnarray}
The last equation shows $a_{0},b_{0},c_{0},d_{0} $ are functions of $t$ only.
Similarly, from  \begin{equation}\label{tt3} T_{t}+T~V=V^{[1]}~T,
\end{equation} comparing the coefficients of $\lambda^{j} ,j =4,3, 2, 1, 0$,it implies
\begin{eqnarray}\label{ttt1}
&&\lambda^{4}: -2 i b_{0}-q^{[1]}d_{1}+q a_{1}=0,\ 2 i c_{0}-2r^{[1]}a_{1}+r d_{1}=0,\nonumber\\
&&\lambda^{3}: -r^{[1]} q^{[1]} a_{1} i-2 q^{[1]} c_{0}+a_{1} r q i+2 r b_{0}=0,\ q a_{0}-q^{[1]}d_{0}=0,\nonumber\\
&&~~~~~~~~r d_{0}-r^{[1]}a_{0}=0,-d_{1} r q i+r^{[1]}q^{[1]}d_{1}i+2qc_{0}-2r^{[1]}b_{0}=0,\nonumber\\
&&\lambda^{2}: a_{0} r q-a_{0} r^{[1]}q^{[1]}=0,\ a_{1} r q^{2}-r^{[1]} {q^{[1]}}^{2} d_{1}-b_{0} r q i +q^{[1]}_{x} d_{1} i-a_{1} q_{x} i- r^{[1]} q^{[1]} b_{0} i=0,\nonumber\\
&&~~~~~~~~c_{0} r q i-{r^{[1]}}^{2} q^{[1]} a_{1}+d_{1} r^{2} q+r^{[1]} q^{[1]} c_{0} i+d_{1} r_{x} i-{r^{[1]}}_{x}a_{1} i=0,\ r^{[1]}q^{[1]}d_{0}-r q d_{0}=0,\nonumber\\
&&\lambda^{1}: {a_{1}}_{t}+{q^{[1]}}_{x} c_{0} i+b_{0} r^{2} q-r^{[1]}{q^{[1]}}^{2} c_{0}+b_{0} r_{x}i=0, \ -r^{[1]}{q^{[1]}}^{2} d_{0}+a_{0} r q^{2}+{q^{[1]}}_{x} d_{0} i-a_{0} q_{x}i=0,\nonumber\\
&&~~~~~~~~d_{0} r_{x}i+d_{0} r^{2} q-{r^{[1]}}^{2}q^{[1]} a_{0}-{r^{[1]}}_{x} a_{0} i=0, \ {d_{1}}_{t}-c_{0} q_{x} i+c_{0} r q^2-{r^{[1]}}^{2}q^{[1]} b_{0}-{r^{[1]}}_{x} b_{0} i=0,\nonumber\\
&&\lambda^{0}: {a_{0}}_{t}={b_{0}}_{t}={c_{0}}_{t}={d_{0}}_{t}=0.
\end{eqnarray}
The last equation shows $a_{0}, b_{0}, c_{0}, d_{0} $ are functions of  $x$ only. So
$a_{0}, b_{0}, c_{0}, d_{0} $ are constants.

In order to get the non-trivial solutions, we present a Darboux transformation under the condition $ a_{0} = 0, d_{0} = 0$.
Based on eq.(\ref{xx1}) and eq.(\ref{ttt1}) and without losing any generality,
let  Darboux matrix $T$ be  the  form  of
\begin{equation}\label{TT}
 T_{1}=T_{1}(\lambda;\lambda_1)=\left( \begin{array}{cc}
a_{1}&0 \\
0 &d_{1}\\
\end{array} \right)\lambda+\left( \begin{array}{cc}
0&b_{0}\\
c_{0} &0\\
\end{array} \right).
\end{equation}
Here $a_{1}, d_{1}$ are undetermined function of ($x$, $t$), which will be expressed by the
eigenfunction associated with $\lambda_1$ in the KN spectral problem.
First of all, we introduce  $n$ eigenfunctions $\psi_j$ as
\begin{eqnarray}
&&\psi_{j}=\left(
\begin{array}{c}\label{jie2}
 \phi_{j}   \\
 \varphi_{j}  \\
\end{array} \right),\ \ j=1,2,....n,\phi_{j}=\phi_{j}(x,t,\lambda_{j}), \
\varphi_{j}=\varphi_{j}(x,t,\lambda_{j}). \label{jie1}
\end{eqnarray}

\noindent {\bf Theorem 1.}{\sl  The  elements of one-fold DT  are  parameterized by the
eigenfunction $\psi_1$ associated with
$\lambda_1$ as
 \begin{eqnarray}
d_{1}=\dfrac{1}{a_{1}}, \ \ a_{1}=-\frac{\varphi_{1}}{\phi_{1}},
\ \ b_{0}=c_{0}=\lambda_{1}, \label{DT1aibi}  \\
%%%%%%%%%%%%%%%%%%%%%%%%%%%%%%%%%%%%%%%
\Leftrightarrow
T_1(\lambda;\lambda_1)=\left(
\begin{array}{cc}
-\lambda \dfrac{\varphi_1}{\phi_1 }& \lambda_1\\
\lambda_1 & -\lambda \dfrac{\phi_1}{\varphi_1 }
\end{array}  \right),  \label{DT1matrix}
\end{eqnarray}
and then the new solutions $q^{[1]}$ and $r^{[1]}$ are given by
\begin{eqnarray}\label{sTT}
q^{[1]}=(\frac{\varphi_{1}}{\phi_{1}})^2q+2i\frac{\varphi_{1}}{\phi_{1}}\lambda_{1},
r^{[1]}=(\frac{\phi_{1}}{\varphi_{1}})^2r-2i\frac{\phi_{1}}{\varphi_{1}}\lambda_{1},
\end{eqnarray}
and the new eigenfunction $\psi_j^{[1]}$ corresponding to $\lambda_j$ is
\begin{equation}
\psi^{[1]}_j=
\left(
\begin{array}{c}
\dfrac{1}{\phi_1}\left|\begin{array}{cc}
-\lambda_j\phi_j & \varphi_j\\
-\lambda_1\phi_1 & \varphi_1
 \end{array}\right| \\ \\
\dfrac{1}{\varphi_1}
\left|\begin{array}{cc}
-\lambda_j\varphi_j & \phi_j\\
-\lambda_1\varphi_1 & \phi_1
 \end{array}\right|
\end{array}
\right).
\end{equation}
}
{\bf Proof.} Note that $(a_{1}d_{1})_{x}=0$ is derived from the eq.(\ref{xx1}), and  then
take $a_{1}=\dfrac{1}{d_{1}}$ in the followings. By transformation eq.(\ref{TT}) and eq.(\ref{xx1}), new solutions are given by
 \begin{eqnarray} \label{TT1}
q^{[1]}=\dfrac{a_{1}}{d_{1}}q-2~i~\dfrac{b_{0}}{d_{1}},r^{[1]}=\dfrac{d_{1}}{a_{1}}q+2~i~\dfrac{c_{0}}{a_{1}}.
\end{eqnarray}
By using a general fact of the DT, i.e.,
 $T_{1}(\lambda;\lambda_{1})|_{\lambda=\lambda_1}\psi_{1}=0$, then eq.(\ref{DT1aibi}) is obtained.
Next, substituting $(a_1,d_1,b_0,c_0)$ given in eq.(\ref{DT1aibi}) back into eq.(\ref{TT1}),
then new solutions are given as eq. (\ref{sTT}). Further, by using the explicit matrix
representation eq.(\ref{DT1matrix}) of $T_1$, then
$\psi^{[1]}_j$ is given by
\begin{equation}
\psi^{[1]}_j=T_1(\lambda;\lambda_1)|_{\lambda=\lambda_j} \psi_j=\left.\left(
\begin{array}{cc}
-\lambda \dfrac{\varphi_1}{\phi_1 }& \lambda_1\\
\lambda_1 & -\lambda \dfrac{\phi_1}{\varphi_1 }
\end{array}  \right)\right|_{\lambda=\lambda_j} \left( \begin{array} {c}
\phi_j\\
\varphi_j
\end{array} \right)=
\left(
\begin{array}{c}
\dfrac{1}{\phi_1}\left|\begin{array}{cc}
-\lambda_j\phi_j & \varphi_j\\
-\lambda_1\phi_1 & \varphi_1
 \end{array}\right| \\ \\
\dfrac{1}{\varphi_1}
\left|\begin{array}{cc}
-\lambda_j\varphi_j & \phi_j\\
-\lambda_1\varphi_1 & \phi_1
 \end{array}\right|
\end{array}
\right).
\end{equation}
Last, a tedious calculation shows that $T_1$ in eq.(\ref{DT1matrix}) and new solutions
indeed satisfy eq.(\ref{tt3}) or (equivalently eq.(\ref{ttt1})). So
KN spectral  problem  is  covariant under transformation $T_1$ in eq.(\ref{DT1matrix}) and
eq.(\ref{sTT}), and thus it is the DT of eq.(\ref{sy1}) and eq.(\ref{sy2}). $\square$

 It is easy to find that $T_1$ is equivalent to the Imai's result(see eq.(7) of ref.\cite{kenji1})
and to the Steudel's result(see eq.(21) of ref.\cite{steduel}). Our derivation is more
transparent, and new solutions $q^{[1]}$ and $r^{[1]}$ can be constructed by the
eigenfunction $\psi_1$, which is a solution of linear partial different equations
eq.(\ref{sy1}) and eq.(\ref{sy2}). This is simpler than Steudel's method to solve nonlinear
Riccati equations. The remaining problem is how to guarantee the validity of the reduction condition,
i.e., $q^{[1]}=-(r^{[1]})^{*}$. We shall solve it at the end of this section by choosing special
eigenfunctions and eigenvalues.\\

2.2 N-fold Darboux transformation for KN system

The key task is to establish the determinant representation of the n-fold DT
for KN system in this subsection. To this purpose, set
$$
\begin{array}{cc}
\textbf{D}=&\left\{\left.\left(\begin{array}{cc}
a& 0\\
0&d
  \end{array} \right)\right| a,d \text{ are complex functions of}\ x\ \text{and}\ t  \right\},\\
%%%%%%%%%%%%%%%%%%%%%%%%%%%%%%%%%%%%%%%%%%
\textbf{A}=&\left\{\left.\left(\begin{array}{cc}
0& b\\
c&0
  \end{array} \right)\right| b,c \text{ are complex functions of}\ x\ \text{and}\ t  \right\},
\end{array}
$$
as ref.\cite{kenji1}.

 According to the form of $T_1$ in eq.(\ref{TT}), the n-fold DT should be the form of \cite{kenji1}
\begin{equation}\label{tnss}T_{n}=T_{n}(\lambda;\lambda_1,\lambda_2, \cdots,\lambda_n)
=\sum_{l=0}^{n}P_{l}\lambda^{l},
\end{equation}
with \begin{eqnarray}
\label{tnsss}
P_{n}=\left( \mbox{\hspace{-0.2cm}}
\begin{array}{cc}
a_{n}\mbox{\hspace{-0.3cm}}&0 \\
0 \mbox{\hspace{-0.3cm}}&d_{n}\\
\end{array}  \mbox{\hspace{-0.2cm}}\right)\in \textbf{D},\ P_{n-1}=\left( \mbox{\hspace{-0.2cm}}
\begin{array}{cc}
0 \mbox{\hspace{-0.3cm}}&b_{n-1} \\
c_{n-1} \mbox{\hspace{-0.3cm}}&0\\
\end{array}  \mbox{\hspace{-0.2cm}}\right)\in \textbf{A},\ P_{l}\in \textbf{D}&\textrm{\small \mbox{\hspace{-0.3cm}}(if $l-n$ is even)}
,\ P_{l}\in \textbf{A}&\textrm{\small \mbox{\hspace{-0.3cm}}(if $l-n$ is odd)}.
\end{eqnarray}
Here $P_{0}$ is a constant matrix, $P_i(1\leq i\leq n)$ is the function of $x$ and $t$.
In particular, $P_0 \in \textbf{D}$ if $n$ is even and $P_0 \in \textbf{A}$ if $n$ is odd, which
leads to the separate discussion on the determinant representation of $T_n$ in the following
by means of its kernel. Specifically, from algebraic equations,
\begin{equation}\label{ttnss}
\psi_{k}^{[n]}=T_{n}(\lambda;\lambda_1,\cdots,\lambda_n)|_{\lambda=\lambda_k}\psi_{k}=\sum_{l=0}^{n}P_{l}\lambda_{k}^{l}\psi_{k}=0,
k=1,2,\cdots,n,
\end{equation}
coefficients $P_i$ is solved by Cramer's rule. Thus we get determinant representation of the $T_n$.

\noindent {\bf Theorem2.} (1)For $n=2k(k=1,2,3,\cdots)$, the n-fold DT of the KN system can be expressed by
\begin{equation}
\label{fss1}T_{n}=T_{n}(\lambda;\lambda_1,\lambda_2,\cdots,\lambda_n)=\left(
\begin{array}{cc}
\dfrac{\widetilde{(T_{n})_{11}}}{W_{n}}& \dfrac{\widetilde{(T_{n})_{12}}}{W_{n}}\\ \\
\dfrac{\widetilde{(T_{n})_{21}}}{\widetilde{W_{n}}}& \dfrac{\widetilde{(T_{n})_{22}}}{\widetilde{W_{n}}}\\
\end{array} \right),
\end{equation}
with
\begin{equation}\label{fsst1}
W_{n}=\begin{vmatrix}
\lambda_{1}^{n}\phi_{1}&\lambda_{1}^{n-1}\varphi_{1}&\lambda_{1}^{n-2}\phi_{1}&\lambda_{1}^{n-3}\varphi_{1}&\ldots&\lambda_{1}^{2}\phi_{1}&\lambda_{1}\varphi_{1}\\
\lambda_{2}^{n}\phi_{2}&\lambda_{2}^{n-1}\varphi_{2}&\lambda_{2}^{n-2}\phi_{2}&\lambda_{2}^{n-3}\varphi_{2}&\ldots&\lambda_{2}^{2}\phi_{2}&\lambda_{2}\varphi_{2}\\
\vdots&\vdots&\vdots&\vdots&\vdots&\vdots&\vdots\\
\lambda_{n}^{n}\phi_{n}&\lambda_{n}^{n-1}\varphi_{n}&\lambda_{n}^{n-2}\phi_{n}&\lambda_{n}^{n-3}\varphi_{n}&\ldots&\lambda_{n}^{2}\phi_{n}&\lambda_{n}\varphi_{n}\nonumber\\
\end{vmatrix},
\end{equation}
\begin{equation}\label{fsst2}
\widetilde{(T_{n})_{11}}=\begin{vmatrix}
\lambda^{n}&0&\lambda^{n-2}&0&\ldots&\lambda^{2}&0&\lambda_{1}\lambda_{2}\ldots\lambda_{n}\\
\lambda_{1}^{n}\phi_{1}&\lambda_{1}^{n-2}\varphi_{1}&\lambda_{1}^{n-2}\phi_{1}&\lambda_{1}^{n-3}\varphi_{1}&\ldots&\lambda_{1}^{2}\phi_{1}&\lambda_{1}\varphi_{1}&\lambda_{1}\lambda_{2}\ldots\lambda_{n}\phi_{1}\\
\lambda_{2}^{n}\phi_{2}&\lambda_{2}^{n-1}\varphi_{2}&\lambda_{2}^{n-2}\phi_{2}&\lambda_{2}^{n-3}\varphi_{2}&\ldots&\lambda_{2}^{2}\phi_{2}&\lambda_{2}\varphi_{2}&\lambda_{1}\lambda_{2}\ldots\lambda_{n}\phi_{2}\\
\vdots&\vdots&\vdots&\vdots&\vdots&\vdots&\vdots&\vdots\\
\lambda_{n}^{n}\phi_{n}&\lambda_{n}^{n-1}\varphi_{n}&\lambda_{n}^{n-2}\phi_{n}&\lambda_{n}^{n-3}\varphi_{n}&\ldots&\lambda_{n}^{2}\phi_{n}&\lambda_{n}\varphi_{n}&\lambda_{1}\lambda_{2}\ldots\lambda_{n}\phi_{1}\nonumber\\
\end{vmatrix},
\end{equation}
\begin{equation}\label{fsst3}
\widetilde{(T_{n})_{12}}=\begin{vmatrix}
0&\lambda^{n-1}&0&\lambda^{n-3}&\ldots&0&\lambda&0\\
\lambda_{1}^{n}\phi_{1}&\lambda_{1}^{n-1}\varphi_{1}&\lambda_{1}^{n-2}\phi_{1}&\lambda_{1}^{n-3}\varphi_{1}&\ldots&\lambda_{1}^{2}\phi_{1}&\lambda_{1}\varphi_{1}&\lambda_{1}\lambda_{2}\ldots\lambda_{n}\phi_{1}\\
\lambda_{2}^{n}\phi_{2}&\lambda_{2}^{n-1}\varphi_{2}&\lambda_{2}^{n-2}\phi_{2}&\lambda_{2}^{n-3}\varphi_{2}&\ldots&\lambda_{2}^{2}\phi_{2}&\lambda_{2}\varphi_{2}&\lambda_{1}\lambda_{2}\ldots\lambda_{n}\phi_{2}\\
\vdots&\vdots&\vdots&\vdots&\vdots&\vdots&\vdots&\vdots\\
\lambda_{n}^{n}\phi_{n}&\lambda_{n}^{n-1}\varphi_{n}&\lambda_{n}^{n-2}\phi_{n}&\lambda_{n}^{n-3}\varphi_{n}&\ldots&\lambda_{n}^{2}\phi_{n}&\lambda_{n}\varphi_{n}&\lambda_{1}\lambda_{2}\ldots\lambda_{n}\phi_{1}\nonumber\\
\end{vmatrix},
\end{equation}
\begin{equation}\label{fsst4}
\widetilde{W_{n}}=\begin{vmatrix}
\lambda_{1}^{n}\varphi_{1}&\lambda_{1}^{n-1}\phi_{1}&\lambda_{1}^{n-2}\varphi_{1}&\lambda_{1}^{n-3}\phi_{1}&\ldots&\lambda_{1}^{2}\varphi_{1}&\lambda_{1}\phi_{1}\\
\lambda_{2}^{n}\varphi_{2}&\lambda_{2}^{n-1}\phi_{2}&\lambda_{2}^{n-2}\varphi_{2}&\lambda_{2}^{n-3}\phi_{2}&\ldots&\lambda_{2}^{2}\varphi_{2}&\lambda_{2}\phi_{2}\\
\vdots&\vdots&\vdots&\vdots&\vdots&\vdots&\vdots\\
\lambda_{n}^{n}\varphi_{n}&\lambda_{n}^{n-1}\phi_{n}&\lambda_{n}^{n-2}\varphi_{n}&\lambda_{n}^{n-3}\phi_{n}&\ldots&\lambda_{n}^{2}\varphi_{n}&\lambda_{n}\phi_{n}\nonumber\\
\end{vmatrix},
\end{equation}
\begin{equation}\label{fsst5}
\widetilde{(T_{n})_{21}}=\begin{vmatrix}
0&\lambda^{n-1}&0&\lambda^{n-3}&\ldots&0&\lambda&0\\
\lambda_{1}^{n}\varphi_{1}&\lambda_{1}^{n-1}\phi_{1}&\lambda_{1}^{n-2}\varphi_{1}&\lambda_{1}^{n-3}\phi_{1}&\ldots&\lambda_{1}^{2}\varphi_{1}&\lambda_{1}\phi_{1}&\lambda_{1}\lambda_{2}\ldots\lambda_{n}\varphi_{1}\\
\lambda_{2}^{n}\varphi_{2}&\lambda_{2}^{n-1}\phi_{2}&\lambda_{2}^{n-2}\varphi_{2}&\lambda_{2}^{n-3}\phi_{2}&\ldots&\lambda_{2}^{2}\varphi_{2}&\lambda_{2}\phi_{2}&\lambda_{1}\lambda_{2}\ldots\lambda_{n}\varphi_{2}\\
\vdots&\vdots&\vdots&\vdots&\vdots&\vdots&\vdots&\vdots\\
\lambda_{n}^{n}\varphi_{n}&\lambda_{n}^{n-1}\phi_{n}&\lambda_{n}^{n-2}\varphi_{n}&\lambda_{n}^{n-3}\phi_{n}&\ldots&\lambda_{n}^{2}\varphi_{n}&\lambda_{n}\phi_{n}&\lambda_{1}\lambda_{2}\ldots\lambda_{n}\varphi_{1}\nonumber\\
\end{vmatrix},
\end{equation}
\begin{equation}\label{fsst6}
\widetilde{(T_{n})_{22}}=\begin{vmatrix}
\lambda^{n}&0&\lambda^{n-2}&0&\ldots&\lambda^{2}&0&\lambda_{1}\lambda_{2}\ldots\lambda_{n}\\
\lambda_{1}^{n}\varphi_{1}&\lambda_{1}^{n-2}\phi_{1}&\lambda_{1}^{n-2}\varphi_{1}&\lambda_{1}^{n-3}\phi_{1}&\ldots&\lambda_{1}^{2}\varphi_{1}&\lambda_{1}\phi_{1}&\lambda_{1}\lambda_{2}\ldots\lambda_{n}\varphi_{1}\\
\lambda_{2}^{n}\varphi_{2}&\lambda_{2}^{n-1}\phi_{2}&\lambda_{2}^{n-2}\varphi_{2}&\lambda_{2}^{n-3}\phi_{2}&\ldots&\lambda_{2}^{2}\varphi_{2}&\lambda_{2}\phi_{2}&\lambda_{1}\lambda_{2}\ldots\lambda_{n}\varphi_{2}\\
\vdots&\vdots&\vdots&\vdots&\vdots&\vdots&\vdots&\vdots\\
\lambda_{n}^{n}\varphi_{n}&\lambda_{n}^{n-1}\phi_{n}&\lambda_{n}^{n-2}\varphi_{n}&\lambda_{n}^{n-3}\phi_{n}&\ldots&\lambda_{n}^{2}\varphi_{n}&\lambda_{n}\phi_{n}&\lambda_{1}\lambda_{2}\ldots\lambda_{n}\varphi_{1}\nonumber\\
\end{vmatrix}.
\end{equation}
(2)For $n=2k+1(k=1,2,3,\cdots)$, then
\begin{equation}\label{fss2}
T_{n}=T_{n}(\lambda;\lambda_1,\lambda_2,\cdots,\lambda_n)=\left(
\begin{array}{cc}
\dfrac{\widehat{(T_{n})_{11}}}{Q_{n}}& \dfrac{\widehat{(T_{n})_{12}}}{Q_{n}}\\ \\
\dfrac{\widehat{(T_{n})_{21}}}{\widehat{Q_{n}}}& \dfrac{\widehat{(T_{n})_{22}}}{\widehat{Q_{n}}}\\
\end{array} \right),
\end{equation}
with
\begin{equation}\label{fsstt1}
Q_{n}=\begin{vmatrix}
\lambda_{1}^{n}\phi_{1}&\lambda_{1}^{n-1}\varphi_{1}&\lambda_{1}^{n-2}\phi_{1}&\lambda_{1}^{n-3}\varphi_{1}&\ldots&\lambda_{1}^{3}\phi_{1}&\lambda_{1}^{2}\varphi_{1}&\lambda_{1}\phi_{1}\\
\lambda_{2}^{n}\phi_{2}&\lambda_{2}^{n-1}\varphi_{2}&\lambda_{2}^{n-2}\phi_{2}&\lambda_{2}^{n-3}\varphi_{2}&\ldots&\lambda_{2}^{3}\phi_{2}&\lambda_{2}^{2}\varphi_{2}&\lambda_{2}\phi_{2}\\
\vdots&\vdots&\vdots&\vdots&\vdots&\vdots&\vdots&\vdots\\
\lambda_{n}^{n}\phi_{n}&\lambda_{n}^{n-1}\varphi_{n}&\lambda_{n}^{n-2}\phi_{n}&\lambda_{n}^{n-3}\varphi_{n}&\ldots&\lambda_{n}^{3}\phi_{n}&\lambda_{n}^{2}\varphi_{n}&\lambda_{n}\phi_{n}\nonumber\\
\end{vmatrix},
\end{equation}
\begin{equation}\label{fsstt2}
\widehat{(T_{n})_{11}}=\begin{vmatrix}
\lambda^{n}&0&\lambda^{n-2}&0&\ldots&\lambda^{3}&0&\lambda&0\\
\lambda_{1}^{n}\phi_{1}&\lambda_{1}^{n-1}\varphi_{1}&\lambda_{1}^{n-2}\phi_{1}&\lambda_{1}^{n-3}\varphi_{1}&\ldots&\lambda_{1}^{3}\phi_{1}&\lambda_{1}^{2}\varphi_{1}&\lambda_{1}\phi_{1}&-\lambda_{1}\lambda_{2}\ldots\lambda_{n}\varphi_{1}\\
\lambda_{2}^{n}\phi_{2}&\lambda_{2}^{n-1}\varphi_{2}&\lambda_{2}^{n-2}\phi_{2}&\lambda_{2}^{n-3}\varphi_{2}&\ldots&\lambda_{2}^{3}\phi_{2}&\lambda_{2}^{2}\varphi_{2}&\lambda_{2}\phi_{2}&-\lambda_{1}\lambda_{2}\ldots\lambda_{n}\varphi_{2}\\
\vdots&\vdots&\vdots&\vdots&\vdots&\vdots&\vdots&\vdots&\vdots\\
\lambda_{n}^{n}\phi_{n}&\lambda_{n}^{n-1}\varphi_{n}&\lambda_{n}^{n-2}\phi_{n}&\lambda_{n}^{n-3}\varphi_{n}&\ldots&\lambda_{n}^{3}\phi_{n}&\lambda_{n}^{2}\varphi_{n}&\lambda_{n}\phi_{n}&-\lambda_{1}\lambda_{2}\ldots\lambda_{n}\varphi_{n}\nonumber\\
\end{vmatrix},
\end{equation}
\begin{equation}\label{fsstt3}
\widehat{(T_{n})_{12}}=\begin{vmatrix}
0&\lambda^{n-1}&0&\lambda^{n-3}&...&0&\lambda^{2}&0&-\lambda_{1}\lambda_{2}\ldots\lambda_{n}\\
\lambda_{1}^{n}\phi_{1}&\lambda_{1}^{n-1}\varphi_{1}&\lambda_{1}^{n-2}\phi_{1}&\lambda_{1}^{n-3}\varphi_{1}&\ldots&\lambda_{1}^{3}\phi_{1}&\lambda_{1}^{2}\varphi_{1}&\lambda_{1}\phi_{1}&-\lambda_{1}\lambda_{2}\ldots\lambda_{n}\varphi_{1}\\
\lambda_{2}^{n}\phi_{2}&\lambda_{2}^{n-1}\varphi_{2}&\lambda_{2}^{n-2}\phi_{2}&\lambda_{2}^{n-3}\varphi_{2}&\ldots&\lambda_{2}^{3}\phi_{2}&\lambda_{2}^{2}\varphi_{2}&\lambda_{2}\phi_{2}&-\lambda_{1}\lambda_{2}\ldots\lambda_{n}\varphi_{2}\\
\vdots&\vdots&\vdots&\vdots&\vdots&\vdots&\vdots&\vdots&\vdots\\
\lambda_{n}^{n}\phi_{n}&\lambda_{n}^{n-1}\varphi_{n}&\lambda_{n}^{n-2}\phi_{n}&\lambda_{n}^{n-3}\varphi_{n}&\ldots&\lambda_{n}^{3}\phi_{n}&\lambda_{n}^{2}\varphi_{n}&\lambda_{n}\phi_{n}&-\lambda_{1}\lambda_{2}\ldots\lambda_{n}\varphi_{n}\nonumber\\
\end{vmatrix},
\end{equation}
\begin{equation}\label{fsstt4}
\widehat{Q_{n}}=\begin{vmatrix}
\lambda_{1}^{n}\varphi_{1}&\lambda_{1}^{n-1}\phi_{1}&\lambda_{1}^{n-2}\varphi_{1}&\lambda_{1}^{n-3}\phi_{1}&\ldots&\lambda_{1}^{3}\varphi_{1}&\lambda_{1}^{2}\phi_{1}&\lambda_{1}\varphi_{1}\\
\lambda_{2}^{n}\varphi_{2}&\lambda_{2}^{n-1}\phi_{2}&\lambda_{2}^{n-2}\varphi_{2}&\lambda_{2}^{n-3}\phi_{2}&\ldots&\lambda_{2}^{3}\varphi_{2}&\lambda_{2}^{2}\phi_{2}&\lambda_{2}\varphi_{2}\\
\vdots&\vdots&\vdots&\vdots&\vdots&\vdots&\vdots&\vdots\\
\lambda_{n}^{n}\varphi_{n}&\lambda_{n}^{n-1}\phi_{n}&\lambda_{n}^{n-2}\varphi_{n}&\lambda_{n}^{n-3}\phi_{n}&\ldots&\lambda_{n}^{3}\varphi_{n}&\lambda_{n}^{2}\phi_{n}&\lambda_{n}\varphi_{n}\nonumber\\
\end{vmatrix},
\end{equation}
\begin{equation}\label{fsstt5}
\widehat{(T_{n})_{21}}=\begin{vmatrix}
0&\lambda^{n-1}&0&\lambda^{n-3}&...&0&\lambda^{2}&0&-\lambda_{1}\lambda_{2}\ldots\lambda_{n}\\
\lambda_{1}^{n}\varphi_{1}&\lambda_{1}^{n-1}\phi_{1}&\lambda_{1}^{n-2}\varphi_{1}&\lambda_{1}^{n-3}\phi_{1}&\ldots&\lambda_{1}^{3}\varphi_{1}&\lambda_{1}^{2}\phi_{1}&\lambda_{1}\varphi_{1}&-\lambda_{1}\lambda_{2}\ldots\lambda_{n}\phi_{1}\\
\lambda_{2}^{n}\varphi_{2}&\lambda_{2}^{n-1}\phi_{2}&\lambda_{2}^{n-2}\varphi_{2}&\lambda_{2}^{n-3}\phi_{2}&\ldots&\lambda_{2}^{3}\varphi_{2}&\lambda_{2}^{2}\phi_{2}&\lambda_{2}\varphi_{2}&-\lambda_{1}\lambda_{2}\ldots\lambda_{n}\phi_{2}\\
\vdots&\vdots&\vdots&\vdots&\vdots&\vdots&\vdots&\vdots&\vdots\\
\lambda_{n}^{n}\varphi_{n}&\lambda_{n}^{n-1}\phi_{n}&\lambda_{n}^{n-2}\varphi_{n}&\lambda_{n}^{n-3}\phi_{n}&\ldots&\lambda_{n}^{3}\varphi_{n}&\lambda_{n}^{2}\phi_{n}&\lambda_{n}\varphi_{n}&-\lambda_{1}\lambda_{2}\ldots\lambda_{n}\phi_{n}\nonumber\\
\end{vmatrix},
\end{equation}
\begin{equation}\label{fsstt6}
\widehat{(T_{n})_{22}}=\begin{vmatrix}
\lambda^{n}&0&\lambda^{n-2}&0&\ldots&\lambda^{3}&0&\lambda&0\\
\lambda_{1}^{n}\varphi_{1}&\lambda_{1}^{n-1}\phi_{1}&\lambda_{1}^{n-2}\varphi_{1}&\lambda_{1}^{n-3}\phi_{1}&\ldots&\lambda_{1}^{3}\varphi_{1}&\lambda_{1}^{2}\phi_{1}&\lambda_{1}\varphi_{1}&-\lambda_{1}\lambda_{2}\ldots\lambda_{n}\phi_{1}\\
\lambda_{2}^{n}\varphi_{2}&\lambda_{2}^{n-1}\phi_{2}&\lambda_{2}^{n-2}\varphi_{2}&\lambda_{2}^{n-3}\phi_{2}&\ldots&\lambda_{2}^{3}\varphi_{2}&\lambda_{2}^{2}\phi_{2}&\lambda_{2}\varphi_{2}&-\lambda_{1}\lambda_{2}\ldots\lambda_{n}\phi_{2}\\
\vdots&\vdots&\vdots&\vdots&\vdots&\vdots&\vdots&\vdots&\vdots\\
\lambda_{n}^{n}\varphi_{n}&\lambda_{n}^{n-1}\phi_{n}&\lambda_{n}^{n-2}\varphi_{n}&\lambda_{n}^{n-3}\phi_{n}&\ldots&\lambda_{n}^{3}\varphi_{n}&\lambda_{n}^{2}\phi_{n}&\lambda_{n}\varphi_{n}&-\lambda_{1}\lambda_{2}\ldots\lambda_{n}\phi_{n}\\
\end{vmatrix}.
\end{equation}

Next, we consider the transformed new solutions ($q^{[n]},r^{[n]}$)of KN system corresponding to
the n-fold DT. Under covariant requirement of spectral problem of the KN system, the transformed
form should be
\begin{equation}\label{nsys11}
\pa_{x}\psi^{[n]}=(J\lambda^2+Q^{[n]}\lambda)\psi=U^{[n]}\psi,
\end{equation}
with
\begin{equation}\label{nfj1}
     \psi=\left( \begin{array}{c}
          \phi   \\
          \varphi \\
      \end{array} \right),
    \quad J= \left( \begin{array}{cc}
       i &0 \\
       0 &-i\\
     \end{array} \right),
  \quad Q^{[n]}=\left( \begin{array}{cc}
    0 &q^{[n]} \\
    r^{[n]} &0\\
\end{array} \right),
\end{equation}
and then  \begin{equation}\label{ntt2} {T_{n}}_{x}+T_{n}~U=U^{[n]}~T_{n}.
\end{equation}
Substituting $T_n$ given by eq.(\ref{tnss}) into eq.(\ref{ntt2}),and then comparing the
coefficients of $\lambda^{n+1}$, it yields
\begin{eqnarray}\label{ntt3}
&&q^{[n]}=\dfrac{a_{n}}{d_{n}}q-2i\dfrac{b_{n-1}}{d_{n}},
\ \ r^{[n]}=\dfrac{d_{n}}{a_{n}}r+2i\dfrac{c_{n-1}}{a_{n}}.
\end{eqnarray}
Furthermore, taking $a_{n},d_{n},b_{n-1},c_{n-1}$
which are obtained from eq.(\ref{fss1}) for $n=2k$ and
from eq.(\ref{fss2}) for $n=2k+1$, into (\ref{ntt3}), then new solutions ($q^{[n]},r^{[n]}$)
are given by
\begin{eqnarray}\label{ntt4}
&&q^{[n]}=\dfrac{\Omega_{11}^{2}}{\Omega_{21}^{2}}q+2i\dfrac{\Omega_{11}\Omega_{12}}{\Omega_{21}^{2}}, \ \ r^{[n]}=\dfrac{\Omega_{21}^{2}}
{\Omega_{11}^{2}}r-2i\dfrac{\Omega_{21}\Omega_{22}}{\Omega_{11}^{2}}.
\end{eqnarray}
Here, (1)for $n=2k$,
\begin{equation}\label{ntt5}
\Omega_{11}=\begin{vmatrix}
\lambda_{1}^{n-1}\varphi_{1}&\lambda_{1}^{n-2}\phi_{1}&\lambda_{1}^{n-3}\varphi_{1}&\ldots&\lambda_{1}\varphi_{1}&\phi_{1}\\
\lambda_{2}^{n-1}\varphi_{2}&\lambda_{2}^{n-2}\phi_{2}&\lambda_{2}^{n-3}\varphi_{2}&\ldots&\lambda_{2}\varphi_{2}&\phi_{2}\\
\vdots&\vdots&\vdots&\vdots&\vdots&\vdots\\
\lambda_{n}^{n-1}\varphi_{n}&\lambda_{n}^{n-2}\phi_{n}&\lambda_{n}^{n-3}\varphi_{n}&\ldots&\lambda_{n}\varphi_{n}&\phi_{n}\\
\end{vmatrix},
\end{equation}
\begin{equation*}
\Omega_{12}=\begin{vmatrix}
\lambda_{1}^{n}\phi_{1}&\lambda_{1}^{n-2}\phi_{1}&\lambda_{1}^{n-3}\varphi_{1}&\ldots&\lambda_{1}\varphi_{1}&\phi_{1}\\
\lambda_{2}^{n}\phi_{2}&\lambda_{2}^{n-2}\phi_{2}&\lambda_{2}^{n-3}\varphi_{2}&\ldots&\lambda_{2}\varphi_{2}&\phi_{2}\\
\vdots&\vdots&\vdots&\vdots&\vdots&\vdots\\
\lambda_{n}^{n}\phi_{n}&\lambda_{n}^{n-2}\phi_{n}&\lambda_{n}^{n-3}\varphi_{n}&\ldots&\lambda_{n}\varphi_{n}&\phi_{n}\\
\end{vmatrix},
\end{equation*}
\begin{equation*}
\Omega_{21}=\begin{vmatrix}
\lambda_{1}^{n-1}\phi_{1}&\lambda_{1}^{n-2}\varphi_{1}&\lambda_{1}^{n-3}\phi_{1}&\ldots&\lambda_{1}\phi_{1}&\varphi_{1}\\
\lambda_{2}^{n-1}\phi_{2}&\lambda_{2}^{n-2}\varphi_{2}&\lambda_{2}^{n-3}\phi_{2}&\ldots&\lambda_{2}\phi_{2}&\varphi_{2}\\
\vdots&\vdots&\vdots&\vdots&\vdots&\vdots\\
\lambda_{n}^{n-1}\phi_{n}&\lambda_{n}^{n-2}\varphi_{n}&\lambda_{n}^{n-3}\phi_{n}&\ldots&\lambda_{n}\phi_{n}&\varphi_{n}\\
\end{vmatrix},
\end{equation*}
\begin{equation*}
\Omega_{22}=\begin{vmatrix}
\lambda_{1}^{n}\varphi_{1}&\lambda_{1}^{n-2}\varphi_{1}&\lambda_{1}^{n-3}\phi_{1}&\ldots&\lambda_{1}\phi_{1}&\varphi_{1}\\
\lambda_{1}^{n}\varphi_{1}&\lambda_{2}^{n-2}\varphi_{2}&\lambda_{2}^{n-3}\phi_{2}&\ldots&\lambda_{2}\phi_{2}&\varphi_{2}\\
\vdots&\vdots&\vdots&\vdots&\vdots&\vdots\\
\lambda_{n}^{n}\varphi_{n}&\lambda_{n}^{n-2}\varphi_{n}&\lambda_{n}^{n-3}\phi_{n}&\ldots&\lambda_{n}\phi_{n}&\varphi_{n}\\
\end{vmatrix};
\end{equation*}
(2) for $n=2k+1$,
\begin{equation}\label{ntt6}
\Omega_{11}=\begin{vmatrix}
\lambda_{1}^{n-1}\varphi_{1}&\lambda_{1}^{n-2}\phi_{1}&\lambda_{1}^{n-3}\varphi_{1}&\ldots&\lambda_{1}\phi_{1}&\varphi_{1}\\
\lambda_{2}^{n-1}\varphi_{2}&\lambda_{2}^{n-2}\phi_{2}&\lambda_{2}^{n-3}\varphi_{2}&\ldots&\lambda_{2}\phi_{2}&\varphi_{2}\\
\vdots&\vdots&\vdots&\vdots&\vdots&\vdots\\
\lambda_{n}^{n-1}\varphi_{n}&\lambda_{n}^{n-2}\phi_{n}&\lambda_{n}^{n-3}\varphi_{n}&\ldots&\lambda_{n}\phi_{n}&\varphi_{n}\\
\end{vmatrix},
\end{equation}
\begin{equation*}
\Omega_{12}=\begin{vmatrix}
\lambda_{1}^{n}\phi_{1}&\lambda_{1}^{n-2}\phi_{1}&\lambda_{1}^{n-3}\varphi_{1}&\ldots&\lambda_{1}\phi_{1}&\varphi_{1}\\
\lambda_{2}^{n}\phi_{2}&\lambda_{2}^{n-2}\phi_{2}&\lambda_{2}^{n-3}\varphi_{2}&\ldots&\lambda_{2}\phi_{2}&\varphi_{2}\\
\vdots&\vdots&\vdots&\vdots&\vdots&\vdots\\
\lambda_{n}^{n}\phi_{n}&\lambda_{n}^{n-2}\phi_{n}&\lambda_{n}^{n-3}\varphi_{n}&\ldots&\lambda_{n}\phi_{n}&\varphi_{n}\\
\end{vmatrix},
\end{equation*}
\begin{equation*}
\Omega_{21}=\begin{vmatrix}
\lambda_{1}^{n-1}\phi_{1}&\lambda_{1}^{n-2}\varphi_{1}&\lambda_{1}^{n-3}\phi_{1}&\ldots&\lambda_{1}\varphi_{1}&\phi_{1}\\
\lambda_{2}^{n-1}\phi_{2}&\lambda_{2}^{n-2}\varphi_{2}&\lambda_{2}^{n-3}\phi_{2}&\ldots&\lambda_{2}\varphi_{2}&\phi_{2}\\
\vdots&\vdots&\vdots&\vdots&\vdots&\vdots\\
\lambda_{n}^{n-1}\phi_{n}&\lambda_{n}^{n-2}\varphi_{n}&\lambda_{n}^{n-3}\phi_{n}&\ldots&\lambda_{n}\varphi_{n}&\phi_{n}\\
\end{vmatrix},
\end{equation*}
\begin{equation*}
\Omega_{22}=\begin{vmatrix}
\lambda_{1}^{n}\varphi_{1}&\lambda_{1}^{n-2}\varphi_{1}&\lambda_{1}^{n-3}\phi_{1}&\ldots&\lambda_{1}\varphi_{1}&\phi_{1}\\
\lambda_{1}^{n}\varphi_{1}&\lambda_{2}^{n-2}\varphi_{2}&\lambda_{2}^{n-3}\phi_{2}&\ldots&\lambda_{2}\varphi_{2}&\phi_{2}\\
\vdots&\vdots&\vdots&\vdots&\vdots&\vdots\\
\lambda_{n}^{n}\varphi_{n}&\lambda_{n}^{n-2}\varphi_{n}&\lambda_{n}^{n-3}\phi_{n}&\ldots&\lambda_{n}\varphi_{n}&\phi_{n}\\
\end{vmatrix}.
\end{equation*}

We are now in a position to consider the reduction of the DT of the KN system so that
$q^{[n]}=-(r^{[n]})^*$, then the DT of the DNLS is given. Under the reduction condition $q=-r^*$,
the eigenfunction $\psi_k=\left( \begin{array}{c}
\phi_k\\
\varphi_k
\end{array} \right)$ associated with eigenvalue $\lambda_k$ has following properties\cite{kenji1},
\begin{description}
\item[(i)] $\phi_{k}^{\ast}=\varphi_{k}$, $\lambda_{k}=-{\lambda_{k}}^{\ast}$;
\item[(ii)] ${\phi_{k}}^{\ast}=\varphi_{l}, \ {\varphi_{k}}^{\ast}=\phi_{l}$,
${\lambda_{k}}^{\ast}=-\lambda_{l}$, where $k\neq l$.
\end{description}
Notice that the denominator $W_n$ of $q^{[n]}$ is a modulus of a non-zero complex function under
reduction condition, so the new solution $q^{[n]}$ is non-singular.
For the one-fold DT $T_1$, set
\begin{equation}\label{onefoldredu}
 \lambda_1=i\beta_1\text{(a pure imaginary constant),\quad and its eigenfunction \quad}
\psi_1=\left( \begin{array}{c}
\phi_1\\
\phi_1^*
\end{array} \right),
\end{equation}
then $T_1$ in theorem 1 is the DT of the DNLS.
We note that $q^{[1]}=-(r^{[1]})^*$ holds  with the help of eq.(\ref{sTT}), $q=-r^*$ and this special
choice of $\psi_1$. This is an essential distinctness of DT between DNLS and NLS, because one-fold
transformation of AKNS can not preserve the reduction condition to the NLS. Furthermore, for the
two-fold DT, according to above property (ii), set
\begin{equation}\label{twofoldredu1}
\lambda_2=-\lambda_1^*\text{\quad and its eigenfunction }
\psi_2=\left( \begin{array}{c}
\varphi_1^*\\
\phi_1^*
\end{array} \right),
\psi_1=\left( \begin{array}{c}
\phi_1\\
\varphi_1
\end{array} \right) \text{associated with eigenvalue\quad } \lambda_1,
\end{equation}
then $q^{[2]}=-(r^{[2]})^*$ can be verified from eq. (\ref{ntt4})
and $T_2$ given by eq.(\ref{fss1}) is the DT of the DNLS.
Of course,  in order to get $q^{[2]}=-(r^{[2]})^*$ so that $T_2$ becomes also
the DT of the DNLS, we  can also set
\begin{equation}\label{twofoldredu2}
\lambda_l=i\beta_l\text{(pure imaginary) and its eigenfunction\ } \psi_l=\left( \begin{array}{c}
\varphi_l^*\\
\phi_l^*
\end{array} \right),l=1,2.
\end{equation}
There are many choices to guarantee  $q^{[n]}=-(r^{[n]})^*$ for the n-fold DTs when $n> 2$. For example,
setting $n=2k$ and $l=1,3,\dots,2k-1$,then choosing following $k$ distinct
eigenvalues and eigenfunctions in n-fold DTs:
\begin{equation}\label{2nfoldredu}
 \lambda_l \leftrightarrow \psi_l=\left( \begin{array}{c}
\phi_l\\
\varphi_l\end{array} \right)
, \text{and} \lambda_{2l}= -\lambda_{2l-1}^*,\leftrightarrow
\psi_{2l}=\left( \begin{array}{c}
\varphi_{2l-1}^*\\
\phi_{2l-1}^*
\end{array} \right)
\end{equation}
so that $q^{[2k]}=-(r^{[2k]})^*$ in eq.(\ref{ntt4}).
Then $T_{2k}$ with these paired-eigenvalue $\lambda_i$
and paired-eigenfunctions $\psi_i(i=1,3,\dots,2k-1)$ is reduced to the (2k)-fold DT of the DNLS.
Similarly, $T_{2k+1}$ in eq.(\ref{fss2}) can also be reduced to the (2k+1)-fold DT of the DNLS
by choosing one pure imaginary  $\lambda_{2k+1}=i\beta_{2k+1}(\text{pure imaginary})$ and  $k$ paired-eigenvalues
$\lambda_{2l}= -\lambda_{2l-1}^*(l=1,2,\cdots,k)$ with corresponding eigenfunctions according to
properties (i) and (ii).
%%%%%%%%%%%%%%%%%%%%%%%%%%%%%%%%%%%%%%%%%%%%%%%%%%%%%%%%55
\section{Particular solutions}
3.1. Darboux transformations  applied to zero seed

 For $q=r=0$ the equations (\ref{sys11}) and (\ref{sys22}) are solved by
 \begin{equation}\label{jie1}
           \psi_{k}=\left( \begin{array}{c}
               \phi_{k} \\
                \varphi_{k}\\
                \end{array} \right),
 \quad \phi_{k}=\exp(i({\lambda_{k}}^{2} x+2 {\lambda_{k}}^{4}t)), \ \  \quad \varphi_{k}=\exp(-i({\lambda_{k}}^{2} x+2 {\lambda_{k}}^{4}t)).
\end{equation}

Case 1(N = 1). Under the choice eq.(\ref{onefoldredu}),
taking $\psi_1$ in eq.(\ref{jie1}) back into  eq.(\ref{sTT}) with  $\lambda_{1}=i\beta_{1}$,
then one solution of the DNLS  is
\begin{equation}\label{onefoldjiefromvacuum}
 q^{[1]}=-2 \beta_{1}\exp(-2i(-{\beta_{1}}^{2}x+2{\beta_{1}}^{4}t)),
\end{equation}
which  is not a soliton but a periodic solution with a constant amplitude.

Case 2(N=2). Considering
the choice in eq.(\ref{twofoldredu2}) with  $\lambda_{1}=i(l+m)$, $\lambda_{2}=i(l-m)$,
and taking eigenfunctions in eq. (\ref{jie1}) back into $T_2$,the result of the DT of the DNLS
is then simply found from (\ref{ntt4}),
\begin{equation}\label{jie21}
 q^{[2]}=-4lm\dfrac{(m\cos(2G)-il\sin(2G))^{3}}{((m^{2}-l^{2})\cos(2G)^{2}+l^{2})^{2}}\exp(2iF),
\end{equation}
which is a quasi-periodic solution,and  here $F=-l^{2} x+2 l^{4} t+12l^{2} m^{2} t-m^{2}x+2 m^{4}t$,
$G=8 l^{3} m t-2l m x+8l m^{3} t$. Furthermore, considering  the choice in eq.(\ref{twofoldredu1}) with
 $\lambda_{1}=\alpha_{1}+i\beta_{1}$, $\lambda_{2}=-\alpha_{1}+i\beta_{1}$, and using eigenfunctions
in eq. (\ref{jie1}), then the solution of the DNLS generated by two-fold DT  is simply found
from (\ref{ntt4}),
\begin{equation}\label{jie22}
 q^{[2]}=4i\alpha\beta\dfrac{(-i\alpha_{1}\cosh(2\Gamma)+\beta_{1}\sinh(2\Gamma))^{3}}{((-{\alpha_{1}}^{2}-{\beta_{1}}^{2})\cosh(2\Gamma)^{2}+{\beta_{1}}^{2})^{2}}\exp(2ih),
\end{equation}
with $h=-{\beta_{1}}^{2} x+2 {\beta_{1}}^{4} t-12{\alpha_{1}}^{2}{\beta_{1}}^{2} t+{\alpha_{1}}^{2}x+2 {\alpha_{1}}^{4}t$, $\Gamma=-8 \alpha_{1} {\beta_{1}}^{3} t+2\alpha_{1} \beta_{1} x+8 {\alpha_{1}}^{3} \beta_{1} t$.
By letting $ \alpha_{1} \rightarrow 0$ in$(\ref{jie22})$,
it becomes a rational solution
\begin{equation} \label{rational1}
q^{[2]}=4 \beta_{1} \exp{(2 i {\beta_{1}}^{2}(-x +2  {\beta_{1}}^{2}t))}
\dfrac{(4i{\beta_{1}}^{2}(4{\beta_{1}}^{2} t-x)-1)^{3}}{(16{\beta_{1}}^{4}({4\beta_{1}}^{2}t-x)^2+1)^2},
\end{equation}
with an arbitrary real constant $\beta_{1}$. Obviously, the rational solution is a linar soliton,
and its trajectory is defined explicitly by
\begin{equation}\label{rational2} x=4 {\beta_{1}}^{2} t,
 \end{equation}
on $(x-t)$ plane. The  solutions $q^{[1]}$ and $q^{[2]}$ of the DNLS equation are consistent with the results
of ref.\cite{kenji1,steduel} except the rational solution. So the rational solution $q^{[2]}$
in eq.(\ref{rational1})  of the DNLS equation is first found in this paper, which is plotted in Figure 1.
\begin{figure}[htbp]     \centering \mbox{}\hspace{0cm}
      \includegraphics[width=0.25\textwidth]{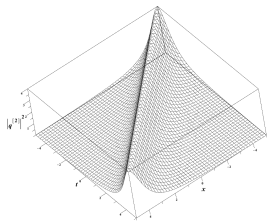}
      \mbox{}\vspace{-0.5cm}\caption{Rational solution $|q^{[2]}|^2$ of the
DNLS with {$\beta_1=0.5$}. }
\end{figure}

3.2. Darboux transformations  applied to non-zero seeds: constant solution and periodic solution.

Set $a$ and  $c$ be two complex constants, and take $c>0$ without loss of generality,
then $ q=c \exp{(i(a x+(-c^2+a) a t))}$ is a periodic solution of the DNLS equation, which will be used as
a seed solution of the DT. Substituting $q=c \exp{(i(a x+(-c^2+a) a t))}$ into the spectral problem eq.(\ref{sys11})
 and eq.(\ref{sys22}), and using the method of separation of variables and the superposition
principle, the eigenfunction $\psi_k$ associated with $\lambda_k$ is given by
\begin{eqnarray}\label{eigenfunfornonzeroseed}
\left(\mbox{\hspace{-0.2cm}} \begin{array}{c}
 \phi_{k}(x,t,\lambda_{k})\\
 \varphi_{k}(x,t,\lambda_{k})\\
\end{array}\mbox{\hspace{-0.2cm}}\right)\mbox{\hspace{-0.2cm}}=\mbox{\hspace{-0.2cm}}\left(\mbox{\hspace{-0.2cm}}\begin{array}{c}
 \varpi1(x,t,\lambda_{k})[1,k]+\varpi2(x,t,\lambda_{k})[1,k]+\varpi1^{\ast}(x,t,-{\lambda_{k}^{\ast})}[2,k]+\varpi2^{\ast}(x,t,-{\lambda_{k}^{\ast})}[2,k]\\
 \varpi1(x,t,\lambda_{k})[2,k]+\varpi2(x,t,\lambda_{k})[2,k]+\varpi1^{\ast}(x,t,-{\lambda_{k}^{\ast})}[1,k]+\varpi2^{\ast}(x,t,-{\lambda_{k}^{\ast})}[1,k]\\
\end{array}\mbox{\hspace{-0.2cm}}\right).
\end{eqnarray}
%%%%%%%%%%%%%%%%%%%%%%
Here
\begin{eqnarray*}
\left(\mbox{\hspace{-0.2cm}}\begin{array}{c}
 \varpi1(x,t,\lambda_{k})[1,k]\\
 \varpi1(x,t,\lambda_{k})[2,k]\\
\end{array}\mbox{\hspace{-0.2cm}}\right)\mbox{\hspace{-0.2cm}}=\mbox{\hspace{-0.2cm}}\left(\mbox{\hspace{-0.2cm}}\begin{array}{c}
\exp(\dfrac{\sqrt{s}( x+2  {\lambda_{k}}^{2} t+(-c^{2}+a) t)}{2}+\dfrac{1}{2}(i(a x+(-c^2+a) a t))) \\
\dfrac{i a -2 i{\lambda_{k}}^{2}+\sqrt{s}}{2{\lambda_{k}} c}\exp(\dfrac{\sqrt{s}(x+2 {\lambda_{k}}^{2} t+(-c^{2}+a) t)}{2}-\dfrac{1}{2}(i(a x+(-c^2+a) a t))) \\
\end{array}\mbox{\hspace{-0.2cm}}\right),\\
\end{eqnarray*}
\begin{eqnarray*}
\left(\mbox{\hspace{-0.2cm}}\begin{array}{c}
 \varpi2(x,t,\lambda_{k})[1,k]\\
 \varpi2(x,t,\lambda_{k})[2,k]\\
\end{array}\mbox{\hspace{-0.2cm}} \right)\mbox{\hspace{-0.2cm}}=\mbox{\hspace{-0.3cm}}
\left(\mbox{\hspace{-0.3cm}}\begin{array}{c}
\exp(-\dfrac{\sqrt{s}( x+2  {\lambda_{k}}^{2} t+(-c^{2}+a) t)}{2}+\dfrac{1}{2}(i(a x+(-c^2+a) a t))) \\
\dfrac{i a -2 i{\lambda_{k}}^{2}-\sqrt{s}}{2{\lambda_{k}} c}\exp(-\dfrac{\sqrt{s}(x+2 {\lambda_{k}}^{2} t+(-c^{2}+a) t)}{2}-\dfrac{1}{2}(i(a x+(-c^2+a) a t)))\\
\end{array}\mbox{\hspace{-0.3cm}}\right)\mbox{\hspace{-0.1cm}},\\
\end{eqnarray*}
\begin{eqnarray*}
\varpi1(x,t,\lambda_{k})=
\left( \begin{array}{c}
 \varpi1(x,t,\lambda_{k})[1,k]\\
 \varpi1(x,t,\lambda_{k})[2,k]\\
\end{array} \right),~~~~~
\varpi2(x,t,\lambda_{k})=
\left( \begin{array}{c}
 \varpi2(x,t,\lambda_{k})[1,k]\\
 \varpi2(x,t,\lambda_{k})[2,k]\\
\end{array} \right),
\end{eqnarray*}
\begin{eqnarray}
&s=-a^{2}-4{\lambda_{k}}^{4}-4{\lambda_{k}}^{2}(c^{2}-a). \nonumber
\end{eqnarray}
Note that
$\varpi1(x,t,\lambda_{k})$ and $\varpi2(x,t,\lambda_{k})$ are two different solutions of
the spectral problem eq.(\ref{sys11}) and eq.(\ref{sys22}),
but we  can only get the  trivial solutions through DT of the DNLS by setting eigenfunction $\psi_k$ be one of them.

    What is more ,we can get  richer solutions by using (\ref{eigenfunfornonzeroseed}).\\
Case 3($N=1$). Under choice in eq. (\ref{onefoldredu}) with
 $\psi_1$ given by eq.(\ref{eigenfunfornonzeroseed}) and $\lambda_{1}=i\beta_{1}$,
the one-fold DT of the DNLS generates
 \begin{equation}\label{q1j1}
 |q^{[1]}|^{2}=c^2-2a+\dfrac{2(2{\beta_{1}}^{2}+a)^{2}-8c^{2}{\beta_{1}}^{2}}{a+2{\beta_{1}}^{2}+2c\beta_{1}\cosh(K(x-2{\beta_{1}}^{2}t+a t-c^2 t))},
\end{equation}
with $K=\sqrt{4c^{2}{\beta_{1}}^{2}-(2{\beta_{1}}^{2}+a)^{2}}$,
according to eq.(\ref{sTT}). By letting $ x \rightarrow {\infty}, t \rightarrow {\infty}$,
 so $|q^{[1]}|^{2} \rightarrow c^2-2a $.  The trajectory is defined implicitly by
 \begin{equation}\label{q1j2} x-2{\beta_{1}}^{2}t+a t-c^2 t =0.
 \end{equation}
The $q^{[1]}$ in eq.(\ref{q1j1}) gives a soliton solutions if
$4c^{2}{\beta_{1}}^{2}-(2{\beta_{1}}^{2}+a)^{2}>0$, and gives a periodic solution
if $4c^{2}{\beta_{1}}^{2}-(2{\beta_{1}}^{2}+a)^{2}<0$. This classification is consistent with
Steudel( see Figure 1 of  ref.\cite{steduel}). Further,  we find that $q^{[1]}$ in
eq.(\ref{q1j1}) can generate  a  dark soliton if $c^2-2 a>(c-2\beta_{1})^{2}$  and a
 bright solitons  if  $c^2-2 a<(c-2\beta_{1})^{2}$. Here
$$|q^{[1]}|^2_{extreme}=(c^2-2a)+
\dfrac{2( (2\beta_1^2+a)^2-4c^2\beta_1^2)}{a+2\beta_1c+\beta_1^2}=(2\beta_1-c)^2.
  $$
Note, $\delta=K^2$ has four roots of $\beta_1$ and $\delta_0=(2\beta_1-c)^2-(c^2-2a)$ has two
roots of $\beta_1$ in general. Combining the conditions of the bright/dark soliton and periodic solutions,
 a complete classification of the different solutions generated by one-fold DT is obtained in
Table 1.  The depth of the dark soliton is $2(-a+2\beta_{1}c-2{\beta_{1}}^{2})$ and the height of
 the bright soliton is $2(a-2\beta_{1}c+2{\beta_{1}}^{2})$. Particularly,  for $a=0$, the seed
solution $q=c$ is a positive constant, and then the one fold DT of the DNLS generates a dark soliton
 under the condition $0<\beta_{1}<c$, the bright soliton under  $-c<\beta_{1}<0$, a periodic solution
under $\beta_1<-c $ and $\beta_{1}>c$. To illustrate the table, Figure 2 is plotted for the case of $c>0$ and $a<0$. Set
$y_1=(c-2\beta_{1})^{2}$, $y_2=4c^{2}{\beta_{1}}^{2}- (2{\beta_{1}}^{2}+a)^{2}=\delta$,
$y_3=c^2-2 a$ with specific parameters $a=-1.5$ $c=0.8$. There are four roots of $y_2$, which are
$(\beta_1)_1>(\beta_1)_2>(\beta_1)_3>(\beta_1)_4$. Note the $(\beta_1)_2$ and
$(\beta_1)_3$ are also the roots of $y_1-y_3=\delta_0$. We can see from Figure 2 that,
$q^{[1]}$ in eq.(\ref{q1j1}) gives the bright soliton  when $\beta_1\in ((\beta_1)_4,
(\beta_1)_3)$ because
$y_2>0$ and $y_1>y_3$, dark soliton when $\beta_1\in ((\beta_1)_2, (\beta_1)_1)$ because
$y_2>0$ and $y_1<y_3$,periodic solutions for others three cases of $\beta_1$ because
$y_2<0$.

\begin{table}[bp]\label{dark-BP}
\caption{{\small  Classification of the solutions $q^{[1]}$  generated by one-fold DT in case 3 according to the intervals
of the eigenvalue $\lambda_1=i\beta_1$.} }
\begin{center}
\begin{tabular}{|c|c|c|c|}\hline
\multicolumn{4}{|c|}{\rule[-0.3cm]{0mm}{0.8cm}\bfseries
  Classification of the solutions  generated by one-fold DT}\\
\hline
zero seed & $c=0$ & $\forall \beta_1\in \mathbb{R}$ &\multicolumn{1}{|c|}{periodic solution}
\\
\hline
\multicolumn{1}{|c|}{\rule[-0.3cm]{0mm}{1cm}constant seed}
& $a=0,c>0$ & $0< \beta_1 <c$& dark solitons\\
  \cline{3-4}
  &{\rule[-0.3cm]{0mm}{1cm}} & $-c< \beta_1<0$
  &bright solitons\\ \cline{3-4}
  &{\rule[-0.3cm]{0mm}{1cm}} & $\beta_1$ belongs to other two intervals &periodic solutions\\
%%%%%%%%%%%%%%%%%%%%%%%%%%%%%%%%%%%%%%%%%%%%%%%%
\hline
  \multicolumn{1}{|c|}{\rule[-0.3cm]{0mm}{1cm}$ \stackrel{\dfrac{c^2}{2}>a}{\text{periodic\ seed}} $ }
& $a>0,c>0$ & $\dfrac{1}{2}c-\dfrac{1}{2}\sqrt{c^{2}-2a}<\beta_{1}<\dfrac{1}{2}c
+\dfrac{1}{2}\sqrt{c^{2}-2a}$ & dark solitons\\
  \cline{3-4}
  &{\rule[-0.3cm]{0mm}{1cm}} &$-\dfrac{1}{2}c-\dfrac{1}{2}\sqrt{c^{2}-2a}<\beta_{1}<-\dfrac{1}{2}c+\dfrac{1}{2}\sqrt{c^{2}-2a}$
  &bright solitons\\ \cline{3-4}
  &{\rule[-0.3cm]{0mm}{1cm}} &$\beta_1$ belongs to other three intervals &periodic solutions\\ \cline{2-4}
  \multicolumn{1}{|c|}{\rule[-0.3cm]{0mm}{1cm}} & $a<0,c>0$ & $-\dfrac{1}{2}c+\dfrac{1}{2}\sqrt{c^{2}-2a}<\beta_{1}<\dfrac{1}{2}c+\dfrac{1}{2}\sqrt{c^{2}-2a}$ & dark solitons \\
  \cline{3-4}
  &{\rule[-0.3cm]{0mm}{1cm}} &$-\dfrac{1}{2}c-\dfrac{1}{2}\sqrt{c^{2}-2a}<\beta_{1}<\dfrac{1}{2}c-\dfrac{1}{2}\sqrt{c^{2}-2a}$
  &bright solitons\\ \cline{3-4}
  &{\rule[-0.3cm]{0mm}{1cm}} & $\beta_1$ belongs to other three intervals
  &periodic solutions\\ \hline
 $\stackrel{ \dfrac{c^2}{2}\leq a}{\text{periodic\ seed}} $ & $a>0,c>0$ & $\forall \beta_1\in \mathbb{R}$ &\multicolumn{1}{|c|}{\rule[-0.75cm]{0mm}{1.5cm}
 ~periodic solutions}  \\ \hline
  \end{tabular}
\end{center}
\end{table}

\begin{figure}[htbp]     \centering \mbox{}\hspace{0cm}
      \includegraphics[width=0.25\textwidth]{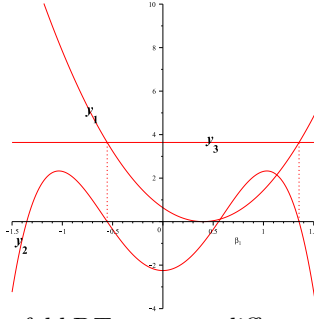}
      \mbox{}\vspace{-0.5cm}\caption{{\footnotesize
           Intervals
of $\beta_1$  in one-fold DT  generate different solutions $q^{[1]}$
(dark soliton, bright soliton and periodic solution) under specific
parameters $a=-1.5,c=0.8$. Here $y_1=(c-2\beta_{1})^{2}$,
$y_2=4c^{2}{\beta_{1}}^{2}- (2{\beta_{1}}^{2}+a)^{2}$, $y_3=c^2-2
a$. There are five intervals of $\beta_1$ divided by the four roots
of $y_2$. From the left to the right, the second interval and the
fourth interval correspond to the bright soliton, dark soliton
respectively. The other three intervals correspond to the periodic
solutions.  }
      }
\end{figure}

%%%%%%%%%%%%%%%%%%%%%%%%
Case 4.$(N=2)$. Under the choice in eq.(\ref{twofoldredu2}) with $\lambda_{1}=i\beta_{1},\lambda_{2}
=i\beta_{2},\beta_{1}\neq \beta_{2}$, the solution of the DNLS equation is generated by two-fold DT
from(\ref{ntt4}) as
\begin{equation}\label{q2j1}
 q^{[2]}=\dfrac{(\beta_{1}{\phi_{1}^{\ast}}\phi_{2}-\beta_{2}\phi_{1}{\phi_{2}^{\ast}})^{2}}{(\beta_{1}\phi_{1}{\phi_{2}^{\ast}}-\beta_{2}{\phi_{1}^{\ast}}\phi_{2})^{2}}q-2\dfrac{({\beta_{1}}^{2}-{\beta_{2}}^{2})\phi_{1}\phi_{2}(\beta_{1}{\phi_{1}^{\ast}}\phi_{2}-\beta_{2}\phi_{1}{\phi_{2}^{\ast}})}{(\beta_{1}\phi_{1}{\phi_{2}^{\ast}}-\beta_{2}{\phi_{1}^{\ast}}\phi_{2})^{2}},
\end{equation}
where  $\phi_{1}$ and $\phi_{2}$ are given by  eq.(\ref{eigenfunfornonzeroseed}).
Similarly, under the choice in eq.(\ref{twofoldredu1}) with one paired eigenvalue
$\lambda_{1}=\alpha_{1}+i\beta_{1}$ and $\lambda_{2}=-\alpha_{1}+i\beta_{1}$,
the two-fold DT eq.(\ref{ntt4}) of the DNLS equation implies  a solution
\begin{equation}\label{q2j2}
 q^{[2]}=\dfrac{(\lambda_{1}\varphi_{1}{\varphi_{1}}^{\ast}-\lambda_{2}\phi_{1}{\phi_{1}}^{\ast})^{2}}{(-\lambda_{2}\varphi_{1}{\varphi_{1}}^{\ast}+\lambda_{1}\phi_{1}{\phi_{1}}^{\ast})^{2}}q+2i\dfrac{({\lambda_{1}}^{2}-{\lambda_{2}}^{2})\phi_{1}{\varphi_{1}}^{\ast}(\lambda_{1}\varphi_{1}{\varphi_{1}}^{\ast}-\lambda_{2}\phi_{1}{\phi_{1}}^{\ast})}{(-\lambda_{2}\varphi_{1}{\varphi_{1}}^{\ast}+\lambda_{1}\phi_{1}{\phi_{1}}^{\ast})^{2}},
\end{equation}
with $\phi_{1}$ and $\varphi_{1}$ given by eq.(\ref{eigenfunfornonzeroseed}).
Two concrete examples of eq.(\ref{q2j2}) are given
below.\\
(a)For simplicity, let $a=2{\alpha_{1}}^{2}-2{\beta_{1}}^{2}+c^{2}$ so that
$\text{Im}(-a^{2}-4{\lambda_{1}}^{4}-4{\lambda_{1}}^{2}(c^{2}-a))=0$, then
\begin{eqnarray}\label{q2j3}
|q^{[2]}|^{2}=-16\alpha_{1}\beta_{1}\dfrac{w1\cosh(f1)\cos(f2)+w2sinh(f1)\sin(f2)+w3}{w4\cosh(f1)\cos(f2)+w5 sinh(f1)\sin(f2)+w6\cos(2 f2)+w7\cosh(2 f1)+w8}+c^2,
\end{eqnarray}
\begin{eqnarray*}
&&w1=c \alpha_{1}(c^{2}-4{\beta_{1}}^{2})(c^{2}+4{\alpha_{1}}^{2}),\\
&&w2=-c\beta_{1}(c^{2}+4 {\alpha_{1}}^{2})(c^{2}-4{\beta_{1}}^{2}),\\
&&w3=2\alpha_{1}\beta_{1}(c^{2}-4{\beta_{1}}^{2})(4{\alpha_{1}}^{2}+c^2),\\
&&w4=8 c {\alpha_{1}}^{2}\beta_{1}(c^{2}+4{\alpha_{1}}^{2}),\\
&&w5=-8 c \alpha_{1}{\beta_{1}}^{2}(c^{2}-4{\beta_{1}}^{2}),\\
&&w6=c^2{\alpha_{1}}^{2}(c^2+4{\alpha_{1}}^{2})+c^2{\beta_{1}}^{2}(c^2-4{\beta_{1}}^{2}),\\
&&w7=16{\alpha_{1}}^{2}{\beta_{1}}^{2}({\alpha_{1}}^{2}+{\beta_{1}}^{2}),\\
&&w8=c^4({\alpha_{1}}^{2}-{\beta_{1}}^{2})+16{\alpha_{1}}^{2}{\beta_{1}}^{2}({\alpha_{1}}^{2}-{\beta_{1}}^{2})+4 c^{2}({\alpha_{1}}^{2}+{\beta_{1}}^{2})^{2},\\
&&f1=K1(4{\alpha_{1}}^{2}t-4{\beta_{1}}^{2}t+x),\\
&&f2=4 K1 \alpha_{1}\beta_{1} t,\\
&&K1=\sqrt{16{\alpha_{1}}^{2}{\beta_{1}}^{2}-4c^{2}{\alpha_{1}}^{2}+4c^{2}{\beta_{1}}^{2}-c^{4}}.
\end{eqnarray*}
By letting $ x \rightarrow {\infty}, \ t \rightarrow {\infty}$, so $|q^{[2]}|^{2} \rightarrow c^2$,
the trajectory of this solution is defined explicitly by
\begin{equation}\label{q1j4}
x=-4{\alpha_{1}}^{2}t+4{\beta_{1}}^{2}t,
 \end{equation}
from $f_1=0$
if $K1^2>0$, and by
\begin{equation}
t=0
\end{equation}
from $f_2=0$ if $(K1)^2<0$. According to eq.(\ref{q2j3}), we can get the Ma breathers\cite{MaB}(time periodic
breather solution) and the Akhmediev breathers\cite{AkhmedievB} (space periodic breather solution)
solution. In general, the solution in eq.(\ref{q2j3}) evolves periodically along the  straight line
with a certain angle of $x$ axis and $t$ axis. The dynamical evolution of
$|q^{[2]}|^2$ in eq.(\ref{q2j3}) for different parameters are plotted in
 Figure 3, Figure 4 and Figure 5, which give a visual verification of the three cases of
 trajectories.  Inspired by the extensive research of rogue wave \cite{ruderman2,AkhmedievB} for
 the nonlinear Schrodinger equation, a limit procedure\cite{AkhmedievB} is used to construct
 rogue wave of the DNLS equation in the following. By letting $ c \rightarrow -2\beta_{1}$ in
 $(\ref{q2j2})$ with $\text{Im}(-a^{2}-4{\lambda_{1}}^{4}-4{\lambda_{1}}^{2}(c^{2}-a))=0$,
it becomes rogue wave
\begin{equation}\label{rationaljie}
q^{[2]}_{rogue~wave}=\dfrac{r1 r2 r3}{r4 r5}
\end{equation}
\begin{eqnarray*}
&&r1=2 exp(2i(\alpha_1^2+\beta_1^2)(2 t \alpha_1^2+x-2 t \beta_1^2))\nonumber\\
&&r2=\beta_1(16 \beta_1^2 \alpha_1^2 (4 t \alpha_1^2+x)^2+16 \beta_1^4 (4t\beta_1^2-x)^2+8i \beta_1^2 (x+4 t \alpha_1^2-8 t \beta_1^2)+1)\nonumber\\
&&r3=\lefteqn{2(16 \beta_1^2\alpha_1^2 (4 t \alpha_1^2+x)^2+16\beta_1^4 (4 t\beta_1^2-x)^2-8\alpha_1\beta_1(x+4 t \alpha_1^2-8 t \beta_1^2)+1){}}\nonumber\\
&&{}\mbox{\hspace{0.8cm}}\times(-\alpha_1+16\beta_1(\beta_1^4-\alpha_1^4) t-4\beta_1(\alpha_1^2+\beta_1^2) x+16i \alpha_1\beta_1^2(\alpha_1^2+\beta_1^2) t-i\beta_1)\nonumber\\
&&{}\mbox{\hspace{0.8cm}}-(16\beta_1^2\alpha_1^2 (4 t\alpha_1^2+x)^2+16\beta_1^4(4 t \beta_1^2-x)^2+8i\beta_1^2(x+4 t \alpha_1^2-8 t\beta_1^2)+1)\nonumber\\
&&{}\mbox{\hspace{0.8cm}}\times(\alpha_1+16\beta_1(\beta_1^4-\alpha_1^4)t-4\beta_1(\alpha_1^2+\beta_1^2) x+16i\alpha_1\beta_1^2(\alpha_1^2+\beta_1^2)t+\beta_1i)\nonumber\\
&&r4=\alpha_1+16\beta_1(\beta_1^4-\alpha_1^4)t-4\beta_1(\alpha_1^2+\beta_1^2)x+16 i\alpha_1\beta_1^2(\alpha_1^2+\beta_1^2)t+\beta_1i\nonumber\\
&&r5=(-16\beta_1^2\alpha_1^2(4 t\alpha_1^2+x)^2-16\beta_1^4(4 t \beta_1^2-x)^2+8i\beta_1^2(x+4 t\alpha_1^2-8 t\beta_1^2)-1)^2\nonumber\\
\end{eqnarray*}
By letting $ x \rightarrow {\infty}, \ t \rightarrow {\infty}$,
 so $|q^{[2]}_{rogue~wave}|^{2}\rightarrow 4\beta_{1}^2$,
the maximum amplitude of $|q^{[2]}_{rogue~wave}|^{2}$ occurs at $t = 0$ and $x=0$ and is
 equal to $36\beta_{1}^{2}$,and the minimum amplitude of $|q^{[2]}_{rogue~wave}|^{2}$ occurs at
  $t = \pm\dfrac{3}{16\sqrt{3(4\alpha_{1}^{2}+\beta_{1}^{2})}\beta_{1}(\alpha_{1}^{2}+\beta_{1}^{2})}$ and
  $x=\mp\dfrac{9\alpha_{1}^{2}}{4\sqrt{3(4\alpha_{1}^{2}+\beta_{1}^{2})}
  \beta_{1}(\alpha_{1}^{2}+\beta_{1}^{2})}$ and is equal to $0$. Through Figure 9 and Figure 10 of
$|q^{[2]}_{rogue~wave}|^2$,  the main features(such as large
amplitude and local property on (x-t) plane) of the rogue wave are
shown. We have found that $|q^{[2]}|^2$ in eq.(\ref{q2j3}) gives the
same result of $|q^{[2]}_{rogue~wave}|^2$ by taking limit of
$c\rightarrow -2\beta_1$.

(b)When $a=\dfrac{c^2}{2}$, from eq.(\ref{eigenfunfornonzeroseed}),
it is not difficult to find that  there are two sets of collinear
eigenfunctions,
\begin{equation}\label{xiangguan1}
\left( \begin{array}{c}
 \varpi1(x,t,\lambda_{k})[1,k]\\
 \varpi1(x,t,\lambda_{k})[2,k]\\
\end{array} \right) \text{and} \ \
\left( \begin{array}{c}
\varpi2^{\ast}(x,t,-{\lambda_{k}^{\ast})}[2,k]\\
\varpi2^{\ast}(x,t,-{\lambda_{k}^{\ast})}[1,k]\\
\end{array}\right),\\
\end{equation}
\begin{equation}\label{xiangguan2}
\left( \begin{array}{c}
 \varpi2(x,t,\lambda_{k})[1,k]\\
 \varpi2(x,t,\lambda_{k})[2,k]\\
\end{array} \right)\text{and} \ \
\left( \begin{array}{c}
\varpi1^{\ast}(x,t,-{\lambda_{k}^{\ast})}[2,k]\\
\varpi1^{\ast}(x,t,-{\lambda_{k}^{\ast})}[1,k]\\
\end{array} \right).
\end{equation}
Therefore,  the eigenfunction $\psi_k$ associated with $\lambda_k$
 for this case is given by
\begin{eqnarray}\label{eigenfunfornonzeroseed1}
\left(\mbox{\hspace{-0.2cm}} \begin{array}{c}
 \phi_{k}(x,t,\lambda_{k})\\
 \varphi_{k}(x,t,\lambda_{k})\\
\end{array}\mbox{\hspace{-0.2cm}}\right)\mbox{\hspace{-0.2cm}}=\mbox{\hspace{-0.2cm}}\left(\mbox{\hspace{-0.2cm}}\begin{array}{c}
 \varpi1(x,t,\lambda_{k})[1,k]+\varpi1^{\ast}(x,t,-{\lambda_{k}^{\ast})}[2,k]\\
 \varpi1(x,t,\lambda_{k})[2,k]+\varpi1^{\ast}(x,t,-{\lambda_{k}^{\ast})}[1,k]\\
\end{array}\mbox{\hspace{-0.2cm}}\right).
\end{eqnarray}
%%%%%%%%%%%%%%%%%%%%%%
Here
\begin{eqnarray*}
\left(\mbox{\hspace{-0.2cm}}\begin{array}{c}
 \varpi1(x,t,\lambda_{k})[1,k]\\
 \varpi1(x,t,\lambda_{k})[2,k]\\
\end{array}\mbox{\hspace{-0.2cm}}\right)\mbox{\hspace{-0.2cm}}=\mbox{\hspace{-0.2cm}}\left(\mbox{\hspace{-0.2cm}}\begin{array}{c}
\exp(i({\lambda_{k}}^{2}x+2{\lambda_{k}}^{4}t+\dfrac{1}{2}c^2 x-\dfrac{1}{4} c^{4} t)) \\
\dfrac{i c}{2 \lambda_{k}}\exp(i({\lambda_{k}}^{2}x+2{\lambda_{k}}^{4}t)) \\
\end{array}\mbox{\hspace{-0.2cm}}\right).
\end{eqnarray*}
Under the choice in eq.(\ref{twofoldredu1}) with
$\lambda_{1}=\alpha_{1}+i\beta_{1},\lambda_{2}=-\alpha_{1}+i\beta_{1}$,
and the $\psi_1$ given by eq.(\ref{eigenfunfornonzeroseed1}), the
solution $q^{[2]}$ is given simply from eq. (\ref{ntt4}). Figure 6
is plotted for $|q^{[2]}|^2$, which shows the  periodical
evolution along a straight line on $(x-t)$ plane.
\\
Case 5.($N = 4$). According to the choice in eq.(\ref{2nfoldredu}) with two distinct  eigenvalues
$\lambda_1=\alpha_1+i\beta_1, \lambda_3=\alpha_3+i\beta_3$, substituting $\psi_1$ and $\psi_3$
defined by eq. (\ref{eigenfunfornonzeroseed}) into eq.(\ref{ntt4}), then the new solution
$q^{[4]}$ generated by 4-fold DT is given. Its analytical expression is omitted because it is
very complicated. But $|q^{[4]}|^2$ are plotted in Figure 7 and 8 to show the
dynamical evolution on $(x-t)$ plane:
(a) Let $a=2{\alpha_{i}}^{2}-2{\beta_{i}}^{2}+c^{2},i=1,3$, so
that $\text{Im}(-a^{2}-4{\lambda_{i}}^{4}-4{\lambda_{i}}^{2}(c^{2}-a))=0$, then Figure 7 shows
intuitively that two breathers may have parallel trajectories;(b) Two breathers
have an elastic collision so that they can preserve their profiles after interaction, which
is verified in Figure 8.
%%%%%%%%%%%%%%%%%%%%%%%%%%%%%%%
\section{Conclusions}

In this paper, a detailed derivation of the DT from the KN system  and then the determinant representation
of the n-fold case are given in Theorem 1 and Theorem 2. Each element of n-fold DT matrix $T_n$ is expressed by the
determinant of eigenfunctions of the spectral problem in eq.(\ref{sys11}) and
eq.(\ref{sys22}). The determinant representations of the new solution $q^{[n]}$ and $r^{[n]}$
of the KN system are also given in  eq.(\ref{ntt4}). Further more, by the special choice of the
eigenvalue $\lambda_k$ and its eigenfunction $\psi_k$ to construct $T_n$ so that $
q^{[n]}=-(r^{[n]})^*$, then the $T_n $ is also reduced to the n-fold DT of the DNLS equation and
 $q^{[n]}$ is a solution of the DNLS. To illustrate our method, solutions of five specific cases are
 discussed by analytical formulae and figures. In particular, a complete classification of the solutions of
 the DNLS equation generated by one-fold DT is given in Table 1.

 By comparing with known results\cite{kenji1,steduel} of the DT for the DNLS equation,
our results provide following improvements:
\begin{itemize}
\item A detailed derivation of the DT and  the determinant representation of $T_n$.
This representation is useful to compute the soliton surfaces of the DNLS  equation in the future as
we have done for the NLS equation\cite{he}. The rogue wave and rational traveling wave are firstly given about
the DNLS equation. The rational solution  has been used by us in a separate preprint to construct the
rouge wave of the variable coefficient DNLS equation\cite{xuhe1}.
%%%%%%%%%%%%%%%%%%%%%%%%%
\item  A complete  and thorough classification of the solution generated by the one-fold DT.
The bright soliton and dark soliton is also classified, which is not published before. At the same time,
our results show the nonlinear and difficult Riccati equations in ref.\cite{steduel},which are transformed from the linear equations of the spectral problem, and Seahorse functions
are indeed avoidable. Of course, these do not disaffirm the merits of method in ref.\cite{steduel}.
%%%%%%%%%%%%%%%%%%%%%%
\item The general solution eq.(\ref{eigenfunfornonzeroseed}) of the linear partial differential equations
in spectral problem is crucial to get non-trivial solution of the DNLS equation.
\item The solution in eq.(\ref{q2j3}) is a relatively general form of the breather solution of the DNLS,
which can evolve periodically along any straight line on $(x-t)$ plane by choosing different values of
parameters $\alpha_1,\beta_1,c$. It has two well-known reductions: Ma breather going periodically along
 $t$-axis, and Akhmediev breather going periodically along $x$-axis.
\end{itemize}

 At last, we would like to mention the DT\cite{fan} of the DNLSIII. Unlike the DNLS equation,
Fan's results show that the kernel of the one-fold DT of the DNLSIII is two dimensional,and
then support again the necessity of the separate study of the three kinds of derivative
nonlinear Schr\"odinger equation. So we shall consider the determinant representation
of the DT for DNLSII and DNLSIII in the near future. Moreover, we are also interested in the
periodic solutions with a variable amplitude of the DNLS equation.

{\bf Acknowledgments} {\noindent \small  This work is supported by
the NSF of China under Grant No.10971109 and K.C.Wong Magna Fund in
Ningbo University. Jingsong He is also supported by Program for NCET
under Grant No.NCET-08-0515. Shuwei Xu is also supported by the Scientific
Research Foundation of Graduate School of Ningbo University. We thank Prof. Yishen Li(USTC,Hefei, China) for his
useful suggestions on the rogue wave.}

\newpage
 \mbox{}
\begin{figure}[htbp]     \centering
      \includegraphics[width=.2\textwidth]{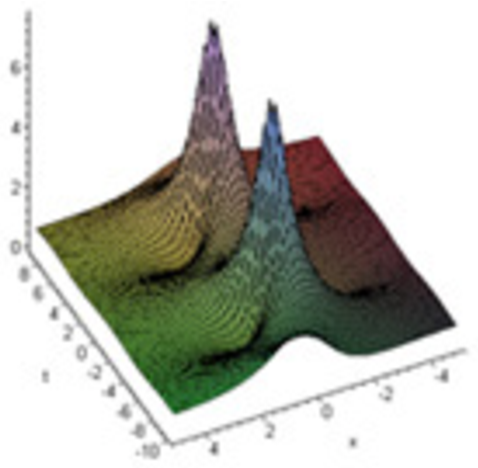} \caption{ The dynamical evolution of
$|q^{[2]}|^2$({\bf time periodic} breather) in eq.(\ref{q2j3}) on
($x-t$) plane with specific parameters
$\alpha_{1}=\beta_{1},\beta_{1}=0.5,c=0.8$. The trajectory is a line
$x=0$. }
\end{figure}
\newpage

\begin{figure}[htbp]     \centering \mbox{}\hspace{0cm}
      \includegraphics[width=.3\textwidth]{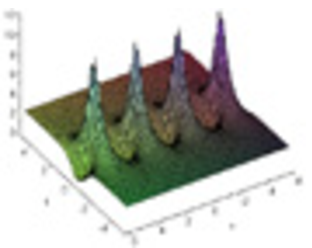}
      \mbox{}\vspace{-0.5cm}\caption{
The
dynamical evolution of $|q^{[2]}|^2$({\bf space periodic} breather) in eq.(\ref{q2j3})
on ($x-t$) plane with specific parameters
$\alpha_{1}=\beta_{1},\beta_{1}=0.5,c=1.5$. The trajectory is a line $t=0$. }
\end{figure}

\begin{figure}[htbp]     \centering \mbox{}\hspace{0cm}
      \includegraphics[width=.3\textwidth]{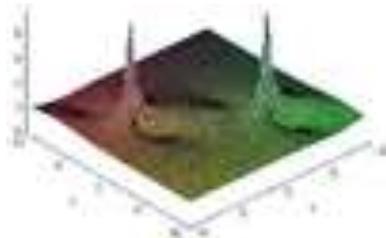}
      \mbox{}\vspace{-0.5cm}\caption{
The dynamical evolution of solution $|q^{[2]}|^2$ in eq.(\ref{q2j3}) for case 4(a).
It evolves periodically along a straight line with
certain angle of $x$ axis and $t$ axis under specific parameters
$\alpha_{1}=0.65,\beta_{1}=0.5,c=0.95$.}
\end{figure}

\begin{figure}[htbp]     \centering \mbox{}\hspace{0cm}
      \includegraphics[width=.3\textwidth]{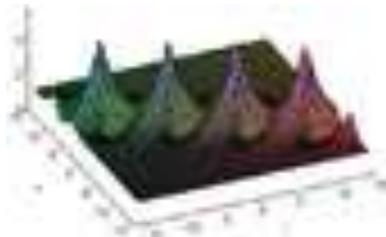}
      \mbox{}\vspace{-0.5cm}\caption{
{The
dynamical evolution of $|q^{[2]}|^2$ in case 4(b) on ($x-t$) plane with specific parameters
$\alpha_{1}=0.5,\beta_{1}=0.35,c=0.85$.} It evolves periodically along a straight line on $(x-t)$
plane.}
\end{figure}

\begin{figure}[htbp]     \centering \mbox{}\hspace{0cm}
      \includegraphics[width=.3\textwidth]{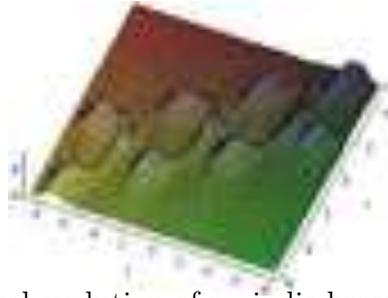}
      \mbox{}\vspace{-0.5cm}\caption{
The dynamical evolution of  periodic breather
solution given by case 5(a) on ($x-t$) plane
 with specific parameters
$\alpha_{1}=0.5,\beta_{1}=0.6,c=0.5,\alpha_{3}=0.6,\beta_{3}=\dfrac{1}{10}\sqrt{47}$.
This picture shows two breathers may parallelly propagate on ($x-t$) plane.}
\end{figure}

\begin{figure}[htbp]     \centering \mbox{}\hspace{0cm}
      \includegraphics[width=.3\textwidth]{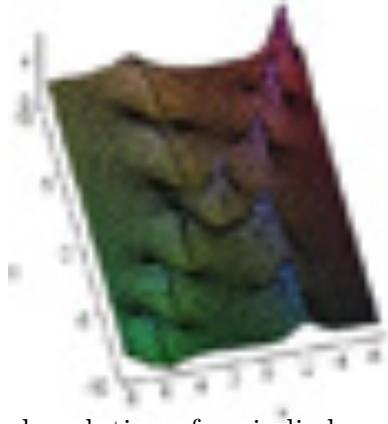}
      \mbox{}\vspace{-0.5cm}\caption{
The
dynamical evolution of periodic breather
solution given by case 5(b) on ($x-t$) plane  with specific parameters
$a=\dfrac{c^2}{2},\alpha_{1}=-0.5,\beta_{1}=0.5,\alpha_{3}=0.6,\beta_{3}=0.5,c=0.95$.
This picture shows the elastic interaction of the two breathers. }
\end{figure}

\begin{figure}[htbp]     \centering \mbox{}\hspace{0cm}
      \includegraphics[width=.3\textwidth]{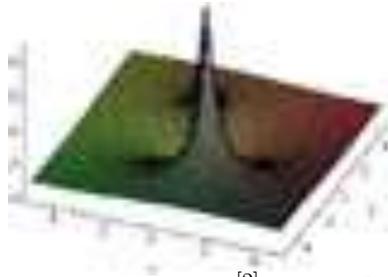}
      \mbox{}\vspace{-0.5cm}\caption{
The
dynamical evolution of $|q^{[2]}_{rogue~wave}|^{2}$ given by eq.(\ref{rationaljie})  on ($x-t$) plane
 with specific parameters
$\alpha_{1}=\dfrac{1}{2},\beta_{1}=\dfrac{1}{2}$.By letting $ x \rightarrow {\infty}, \ t \rightarrow {\infty}$, so $|q^{[2]}_{rogue~wave}|^{2}\rightarrow 1$,
the maximum amplitude of $|q^{[2]}_{rogue~wave}|^{2}$ occurs at $t = 0$ and $x=0$ and is equal to 9,and the minimum amplitude of $|q^{[2]}_{rogue~wave}|^{2}$ occurs at $t = \pm\dfrac{\sqrt{15}}{10}$ and $x=\mp\dfrac{3\sqrt{15}}{10}$ and is equal to $0$. }
\end{figure}

\begin{figure}[htbp]     \centering \mbox{}\hspace{0cm}
      \includegraphics[width=.3\textwidth]{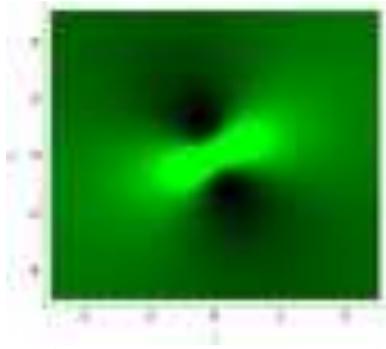}
      \mbox{}\vspace{-0.5cm}\caption{
Contour plot of the wave amplitudes of
$|q^{[2]}_{rogue~wave}|^{2}$ in the ($x-t$) plane is given by eq.(\ref{rationaljie}) for $\alpha_{1}=\dfrac{1}{2},\beta_{1}=\dfrac{1}{2}$. }
\end{figure}

\end{document}